\documentclass[traditabstract]{aa}
\usepackage[english]{babel}

\usepackage[fleqn]{amsmath}
\usepackage{amssymb}
\usepackage{graphicx}
\usepackage{txfonts}
\usepackage{url}

\usepackage{tensind}
\tensordelimiter{ }
\tensorformat{lrc}

\usepackage{color}

\newcommand{\ie}{\textit{i.e.}~}
\newcommand{\eg}{\textit{e.g.}~}

\newcommand{\bF}{\boldsymbol{F}}
\newcommand{\bS}{\boldsymbol{S}}
\newcommand{\bu}{\boldsymbol{u}}
\newcommand{\bx}{\boldsymbol{x}}
\newcommand{\MM}{\mathcal{M}}

\usepackage{natbib}
\bibpunct{(}{)}{;}{a}{}{,} 

\title{\texttt{THC}: a new high-order finite-difference
  high-resolution shock-capturing code for special-relativistic
  hydrodynamics}

\titlerunning{\texttt{THC}, a new high-order FD HRSC
  code for special relativistic hydrodynamics}

\author{David Radice\inst{1} \and Luciano Rezzolla\inst{1,2}}

\institute{Max-Planck-Institut f\"ur Gravitationsphysik, Albert
  Einstein Institut, Potsdam, Germany \and Department of Physics and
  Astronomy, Louisiana State University, Baton Rouge, USA}

\abstract{We present \texttt{THC}: a new high-order
  flux-vector-splitting code for Newtonian and special-relativistic
  hydrodynamics designed for direct numerical simulations of turbulent
  flows. Our code implements a variety of different reconstruction
  algorithms, such as the popular weighted essentially non oscillatory
  and monotonicity-preserving schemes, or the more specialised
  bandwidth-optimised WENO scheme that has been specifically designed
  for the study of compressible turbulence.

  We show the first systematic comparison of these schemes in
  Newtonian physics as well as for special-relativistic flows. In
  particular we will present the results obtained in simulations of
  grid-aligned and oblique shock waves and nonlinear, large-amplitude,
  smooth adiabatic waves. We will also discuss the results obtained in
  classical benchmarks such as the double-Mach shock reflection test
  in Newtonian physics or the linear and nonlinear development of the
  relativistic Kelvin-Helmholtz instability in two and three
  dimensions. Finally, we study the turbulent flow induced by the
  Kelvin-Helmholtz instability and we show that our code is able to
  obtain well-converged velocity spectra, from which we benchmark the
  effective resolution of the different schemes.}

\begin{document}
\maketitle


\section{Introduction}
Numerical relativistic hydrodynamics has come a long way since the
pioneering works by \cite{May66} and \cite{Wilson72} and it is now
playing a central role in modelling of systems involving strong gravity
and/or flows with high Lorentz factors. Example of applications are
relativistic jets, core-collapse supernovae, merger of compact
binaries and the study of gamma-ray bursts [see \cite{Marti03} and
  \cite{Font08} for a complete overview].

In all of these areas the progress has been continuous over the past
few years to the point that relativistic computational fluid dynamics
is starting to provide a realistic description of many
relativistic-astrophysics scenarios [see, \eg
  \cite{Rezzolla:2011}]. Key factors in this progress have been the
switch to more advanced and accurate numerical schemes, and in
particular the adoption of high resolution shock capturing (HRSC)
schemes \citep{Marti91, Schneider90, Banyuls97, Donat98, Aloy99a,
  Baiotti03a} and the progressive inclusion of more ``physics'' for a
more accurate description of the different scenarios. Examples of the
latter are the inclusion of magnetic fields \citep{Koide99,
  DelZanna2003, Gammie03, Komissarov04, Duez05MHD0, Neilsen2005,
  Anton06, Giacomazzo:2007ti} the use of realistic tabulated equations
of state, [see, \eg \cite{Sekiguchi2011}], and the description of
radiative processes \citep{Farris08, Sekiguchi2010, Zanotti2011}.

We expect that both improved physical models and better numerical
techniques will be key elements in the future generation of codes for
relativistic astrophysics. On the one hand it is necessary to take
into account many physical phenomena that are currently oversimplified
and, on the other hand, higher accuracy is necessary to make
quantitative predictions even in the case where simplified models are
used to describe the objects of study. For example, in the case of
inspiralling binary neutron stars, waveforms that are sufficiently
accurate for gravitational-waves templates are just now becoming
available and only in the simple case of polytropic stars
\citep{Baiotti:2009gk,Baiotti:2010,Bernuzzi2011}. Clearly, even higher
accuracy will be required as more realistic equations of state are
considered or better characterisations of the tidal effects are
explored \citep{Baiotti2011,Bernuzzi2012}.

For this reason the development of more accurate numerical tools for
relativistic hydrodynamics is an active and lively field of
research. Most of the effort has been directed towards the development
of high-order finite-volume \citep{tchekhovskoy_2007_wham,
  Duffell2011} and finite-difference \citep{Zhang2006, DelZanna2007}
schemes, but many alternative approaches have been also proposed,
including finite-element methods \citep{Mann1985, meier_1999_mas},
spectral methods \citep{gourgoulhon_1991_seg},
smoothed-particle-hydrodynamics \citep{Siegler00, rosswog_2010_csr}
and discontinuous Galerkin methods \citep{Dumbser2009, Radice2011}.

The use of flux-conservative finite-difference HRSC schemes is
probably the easiest way of increasing the (formal) order of accuracy
of the current generation of numerical codes: finite-difference
schemes are much cheaper then high-order finite-volume codes since
they do not require the solution of multiple Riemann problems at the
interface between different regions \citep{Shu99, Shu01} and they are
free from the complicated averaging and de-averaging procedures of
high-order finite-volume codes [see, \eg
  \cite{tchekhovskoy_2007_wham}].

Here we present a new code, the Templated-Hydrodynamics Code
(\texttt{THC}), developed using the Cactus framework
\citep{Goodale02a}, that follows this approach. \texttt{THC} employs a
state-of-the-art flux-vector splitting scheme: it uses up to
seventh-order reconstruction in characteristics fields and the Roe
flux split with a novel entropy-fix prescription. The ``templated''
aspect reflects the fact that the code design is based on a modern
C$++$ paradigm called template metaprogramming, in which part of the
code is generated at compile time.  Using this particular
  programming technique it is possible to construct object-oriented,
  highly modular, codes without the extra computational costs
  associated with classical polymorphism, because, in the templated
  case, polymorphism is resolved at compile time allowing the compiler
  to inline all the relevant function calls, see \eg
  \citet{yang_2000_oon}. Among the different reconstruction schemes
that we implemented are the classical monotonicity-preserving (MP) MP5
scheme \citep{suresh_1997_amp, mignone_2010_hoc}, the weighted
essentially non oscillatory (WENO) schemes WENO5 and WENO7
\citep{Liu1994, Jiang1996, Shu97} and two bandwidth-optimized WENO
schemes: WENO3B and WENO4B \citep{martin_2006_bow, taylor_2007_one},
designed for direct simulations of compressible turbulence (we recall
that the number associated to the different methods indicates the
putative order of accuracy).

By far the largest motivation behind the development of \texttt{THC}
is the study of the statistical properties of relativistic turbulence
and the determination of possible new and non-classical features. In a
recent paper (Radice \& Rezzolla, in prep.), we have presented the results of
direct numerical simulations of driven turbulence in an
ultrarelativistic hot plasma obtained using \texttt{THC}. More
specifically, we have studied the statistical properties of flows with
average Mach number ranging from $\sim 0.4$ to $\sim 1.7$ and with
average Lorentz factors up to $\sim 1.7$, finding that flow
quantities, such as the energy density or the local Lorentz factor,
show large spatial variance even in the subsonic case as
compressibility is enhanced by relativistic effects. We have also
uncovered that the velocity field is highly intermittent, but its
power-spectrum is found to be in good agreement with the predictions
of the classical theory of Kolmogorov. Overall, the results presented
in Radice \& Rezzolla in prep.~indicate that relativistic effects are able to
enhance significantly the intermittency of the flow and affect the
high-order statistics of the velocity field, while leaving unchanged
the low-order statistics, which appear to be universal and in good
agreement with the classical Kolmogorov theory.

In this paper we give the details of the algorithms used in
\texttt{THC} and employed in Radice \& Rezzolla in prep., presenting a
systematic comparison between the results obtained using the above
mentioned reconstruction schemes, with emphasis for the application of
these schemes for direct simulations of relativistic turbulence. To
our knowledge this is the first time that such a comparison has been
done in the relativistic case.

The rest of this paper is organised as follows. In Section
\ref{sec:code} we present \texttt{THC} code in more detail: we discuss
the numerical algorithms it uses and recall the equations of Newtonian
and special-relativistic hydrodynamics. The results obtained with our
code in a representative number of test cases for the Newtonian and
special relativistic hydrodynamics are presented in Section
\ref{sec:tests}. In Section \ref{sec:kh3d} we present the application
of our code to the study of the linear and nonlinear development of
the relativistic Kelvin-Helmholtz instability (KHI) in three
dimensions as a nontrivial application for our code and a stringent
test of its accuracy. Finally Section \ref{sec:conclusions} is
dedicated to the summary and the conclusions.

\section{The Templated-Hydrodynamics Code}\label{sec:code} 

We here briefly outline the numerical infrastructure adopted by our
templated-hydrodynamics code and report the formulations of the
equations of Newtonian and special-relativistic hydrodynamics we
actually solve.

\subsection{Finite-differencing high-resolution shock-capturing schemes}

We can only provide here a minimal description of the numerical
schemes used by \texttt{THC} and we refer the interested reader to
\cite{Shu99} for a more detailed description of finite-difference
ENO/WENO HRSC schemes, and to \cite{mignone_2010_hoc} for a detailed
description of the finite-difference MP5 scheme. We also note that an
approach very similar to ours was already adopted by \cite{Zhang2006}
in the construction of the \texttt{RAM} code.

Both in the case of the Newtonian and in that of the
relativistic-hydrodynamic equations, we consider a system of
hyperbolic balance-laws in the form
\begin{equation}\label{eq:claw}
\frac{\partial \bF^{0}(\bu)}{\partial t} +
  \sum_{i=1}^3 \frac{\partial \bF^{i}(\bu)}{\partial x^i} = \bS(\bu)\,,
\end{equation}
where the components of $\boldsymbol{u}$ are the so-called
\emph{primitive variables}, a set of $N$ independent physical
quantities, $N$ being the number of equations composing
(\ref{eq:claw}), while $\boldsymbol{F}^0(\boldsymbol{u})$ is referred
to as the vector of \emph{conserved variables} (or state vector), even
though they are not strictly conserved in the presence of a source
term $\bS(\bu)$.

As customary with grid-based codes, we solve numerically the set of
equations (\ref{eq:claw}) on a computational grid defined in a
Cartesian coordinate system at uniformly distributed spatial positions
\begin{equation}
  \bx_{i,j,k} = (i \Delta^1, j \Delta^2, k \Delta^3)\,, \qquad
  i, j, k \in \mathbb{Z}\,,
\end{equation}
and rewrite it using the method of lines, in a semi-discrete,
dimensionally unsplit form as
\begin{align}
\label{eq:mol}
  \frac{d \bF^0_{i,j,k}}{d t} = 
  \bS_{i,j,k} & +
  \frac{\bF^1_{i-1/2,j,k} - \bF^1_{i+1/2,j,k}}{\Delta^1}  \nonumber \\
  & + \frac{\bF^2_{i,j-1/2,k} - \bF^2_{i,j+1/2,k}}{\Delta^2} 
    + \frac{\bF^3_{i,j,k-1/2} - \bF^3_{i,j,k+1/2}}{\Delta^3},
\end{align}
where $\psi_{i,j,k}$ is the value of a generic quantity, $\psi$, at
$\bx_{i,j,k}$, while $(\bF^1_{i-1/2,j,k} - \bF^1_{i+1/2,j,k} ) /
\Delta^1$ is an high-order, non-oscillatory\footnote{We recall that a
  numerical method is said to be \emph{non-oscillatory} if $N(u^{n+1})
  \leq N(u^{n})$, where $N(u^{n})$ denotes the number of local extrema
  of some discrete representation of the function $u$ at time
  $t=n\Delta t$. An \emph{essentially} non-oscillatory method is a
  method that is non-oscillatory on average only. Note that here the
  non-oscillatory feature is really referred to the reconstructed
  derivative of $u$.}, approximation of $-\partial \bF^1 / \partial
x^1$ at $\bx_{i,j,k}$, to be specified. We use the method of lines to
evolve (\ref{eq:mol}) using a third-order strong stability-preserving
Runge-Kutta (SSP-RK) scheme \citep{gottlieb2009}.

To illustrate how we compute the discrete derivatives in the
right-hand-side of (\ref{eq:mol}) it is useful to make a step back and
consider, first a simpler scalar hyperbolic equation in one dimension,
\ie
\begin{equation}\label{eq:scalar.claw}
  \frac{\partial u}{\partial t} + \frac{\partial f(u)}{\partial x} = 0.
\end{equation}
Introducing a uniform grid, $x_i = i \Delta$, and define, for any function,
$v(x)$ the volume averages
\begin{equation}
\label{eq:average}
  \tilde{v}_i \equiv \frac{1}{\Delta} \int_{x_{i-1/2}}^{x_{i+1/2}} v(x)\, d x.
\end{equation}

A reconstruction operator, $\mathcal{R}$, is a nonlinear operator
yielding an high-order approximation of $v$ at a given point $x$ using
its volume averages, $\tilde{v}_i$. Since $v(x)$ can be discontinuous, we
distinguish two different reconstruction operators, $\mathcal{R}^-$ and
$\mathcal{R}^+$, called the left-biased and right-biased reconstruction
operators, such that
\begin{align}
 & \left[\mathcal{R}^-\left(\{\tilde{v}_i\}\right)\right](x) =
    \lim_{y \to x^{-}} v(y) + \mathcal{O}(\Delta^r)\,, \\
 & \left[\mathcal{R}^+\left(\{\tilde{v}_i\}\right)\right](x) =
    \lim_{y \to x^{+}} v(y) + \mathcal{O}(\Delta^r)\,,
\end{align}
where we have indicated with
$\mathcal{R}^+\left(\{\tilde{v}_i\}\right)$ the notion that
$\mathcal{R}$ is an operator that acts on a set of averages
$\tilde{v}_i$, and where $r$ is the order of the reconstruction
operator $\mathcal{R}$. Hereafter we will use the notation
$v_{i+1/2}^-$ and $v_{i+1/2}^+$ to denote the reconstructed values in
$x_{i+1/2}$ using $\mathcal{R}^-$ and $\mathcal{R}^+$
respectively. Example of such operators are given by the ENO/WENO and
the MP5 algorithms.

In classical WENO schemes, the reconstruction is obtained from a
weighted average of a set of lower-order reconstructions of the
$\tilde{v}_i$'s on a number of overlapping stencils. The weights are
computed using some nonlinear smoothness indicators designed in such a
way that the maximum possible order of accuracy is obtained in the
case of smooth solutions. On the other hand, when discontinuities are
detected, the order is automatically reduced to avoid spurious
oscillations. The bandwidth-optimized WENO schemes differ from the
classical ones because they use a more symmetric stencil and because
their weights are not designed to yield the maximum possible formal
order of accuracy for smooth solutions, but are instead tuned to
minimise the attenuation of high-frequency modes. In other words,
while in the classical WENO case the weights for a smooth function are
chosen to match as many terms as possible in the Taylor expansion of
the target function, in the bandwidth-optimized case the coefficients
are chosen to yield the best possible approximation of the Fourier
coefficients of the function to be reconstructed. The nonlinear
smoothness indicators are also modified to avoid ``over-adaptation''
of the scheme and minimise the amount of numerical dissipation [see
\cite{martin_2006_bow, taylor_2007_one} for details].

The MP5 scheme, on the other hand, is based on a fifth-order
reconstruction combined with a flattening procedure designed to avoid
the creation of artificial extrema in the function to be
reconstructed. The monotonicity-preserving reconstruction has the nice
property that no spurious oscillations can be produced by the
reconstruction, which is not guaranteed for the WENO schemes. On the
other hand the limiting procedure employed by MP5 requires the use of
a series of conditional statements in the code, that are not present
in the WENO case.

The reconstruction operators are the core ingredients of both
finite-volume and finite-difference schemes. In a finite-volume scheme
they are used to compute the left and right state to be used in the
(usually approximate) Riemann solver to compute the fluxes. In a
finite-difference scheme, instead, they are used to compute the above
mentioned non-oscillatory approximation of $\partial f / \partial x$.
Following \cite{Shu88} we introduce a function $h(x)$ and such that
\begin{equation}
\label{eq:h_of_x_1}
  f\big[u(x_i)\big] = \frac{1}{\Delta} \int_{x_{i-1/2}}^{x_{i+1/2}}
  h(\xi)\, d\xi\,,
\end{equation}
that is, the average of $h(x)$ between $x_{i-1/2}$ and $x_{i+1/2}$
corresponds to the value of $f$ at $x_{i}$. Next, we note that
\begin{equation}
\label{eq:h_of_x_2}
  \frac{\partial f}{\partial x}\bigg|_{x_i} =
  \frac{h(x_{i+1/2})-h(x_{i-1/2})}{\Delta}\,,
\end{equation}
where both~\eqref{eq:h_of_x_1} and \eqref{eq:h_of_x_2} are exact
expressions. Hence, by using the usual reconstruction operators
$\mathcal{R}$ of order $r$ to reconstruct $h_{i+1/2}$, one obtains a
corresponding accurate approximation of order $r$ of the derivative
$\partial f / \partial x$ at $x_{i}$. Note that $h$ is never actually
computed at any time during the calculation as we only need the values
of $f$ at the gridpoints, \ie $f\big[u(x_i)\big]$.

In order to ensure the stability of the resulting scheme, one has to take
care to upwind the reconstruction appropriately. Let us first consider
the case in which $f'(u) \equiv \partial f /\partial u > 0$. If we set
\begin{equation}
  \tilde{v}_i = f\big[u(x_i)\big] = \frac{1}{\Delta}
  \int_{x_{i-1/2}}^{x_{i+1/2}} h(\xi)\, d\xi
\end{equation}
and 
\begin{align}
& f_{i+1/2} \equiv v_{i+1/2}^-\,,
&& f_{i-1/2} \equiv v_{i-1/2}^-\,,
\end{align}
then 
\begin{equation}
  \frac{\partial f(u)}{\partial x} = \frac{f_{i+1/2} - f_{i-1/2}}{\Delta}
    + \mathcal{O}(\Delta^r)\,,
\end{equation}
gives the wanted high-order approximation of $\partial f / \partial x$
at $x_i$.

In the more general case, where the sign of $f'(u)$ is undetermined, in
order to compute $f_{i+1/2}$, we have to split $f$ in a right-going,
$f^+$, and a left-going, $f^-$, flux, $f = f^+ + f^-$, and use the
appropriate upwind-biased reconstruction operators separately on both
parts, in order to guarantee the stability of the method.

There are several different ways to perform such a split and in our
code we implemented two of them: the Roe flux-split, \ie
\begin{equation}\label{eq:roe.flux.split}
f = f^{\pm}, \quad \textrm{if } [f'(\bar{u})]_{x_{i+1/2}} \gtrless 0\,,
\end{equation}
where $\bar{u}_{i+1/2} \equiv \tfrac{1}{2} (u_i + u_{i+1})$, and the
Lax-Friedrichs or Rusanov flux-split \citep{Shu97}, \ie
\begin{equation}\label{eq:lax.friedrichs.split}
  f^\pm = f(u) \pm \alpha u, \quad \alpha = \max [f'(u)]\,,
\end{equation}
where the maximum is taken over the stencil of the reconstruction
operator. The Roe flux-split is less dissipative and yields a
computationally less-expensive scheme, since only one reconstruction
is required instead of two, but its use can result in the creation of
entropy-violating shocks in the presence of transonic rarefaction
waves [see, \eg \cite{LeVeque92}], and it is also susceptible to the
carbuncle (or odd-even decoupling) phenomenon \citep{Quirk1994}. To
avoid these drawbacks, we switch from the Roe to the Lax-Friedrichs
flux split when $u$ or $f$ are not monotonic within the reconstruction
stencil.

We note that the condition that we use to switch from the Roe to the
Lax-Friedrichs flux split is weaker than the commonly employed condition
on the sign of $f'(u)$, [see, \eg \cite{LeVeque92}], in the sense that it
results in a more frequent use of the Lax-Friedrichs split with respect
to the usual one. According to our experience this prescription works
very well: it is computationally less expensive to compute with respect
to the standard one, since $u$ and $f$ are already evaluated on the grid,
while $f'(u)$ is not, and it seems to be sufficient to avoid the
carbuncle phenomenon in all the tests that we performed. All the results
that we are going to present in this paper have been obtained using this
Roe-split with this ``entropy fix''.

We now go back to the more general system of equations
(\ref{eq:claw}). The derivatives $\partial \bF^a_{i,j,k} / \partial
x^a$ can be computed following the procedure outlined above on a
component-by-component basis. This approach is commonly adopted in the
case of low-order schemes, but it often results in spurious numerical
oscillations in the high-order (usually higher then second) case. To
avoid this issue the reconstruction should be performed on the local
characteristic variables of the systems. To avoid an excessively
complex notation, let us concentrate on the fluxes in the
$x$-direction; in this case, to reconstruct $\bF^1_{i+1/2,j,k}$ we
introduce the Jacobian matrices
\begin{equation}
\label{eq:jacobians}
  \boldsymbol{A}^\alpha = \frac{\partial \bF^\alpha}{\partial \bu}
    \bigg|_{\bar{\bu}}\,, \quad \alpha = 0, 1,
\end{equation}
where
\begin{equation}
\label{eq:uaverage}
\bar{\bu} \equiv \frac{1}{2} (\bu_{i,j,k} + \bu_{i+1,j,k}) \,,
\end{equation}
is the average state at the point where the reconstruction is to be
performed. We point out that the average~\eqref{eq:uaverage}, and
appearing in (\ref{eq:roe.flux.split}) and (\ref{eq:jacobians}), is
much simpler than the average of $\bu_{i,j,k}$ and $\bu_{i+1,j,k}$
suggested by \cite{Roe81}. \cite{Zhang2006} have checked that the use
of~\eqref{eq:uaverage} in place of the average suggested by
\cite{Roe81} does not influence significantly the quality of the
solution in the case of finite-differencing schemes, even in the
relativistic case.

Hyperbolicity of (\ref{eq:claw}) implies that $\boldsymbol{A}^0$ is
invertible and the generalised eigenvalue problem
\begin{equation}
[\boldsymbol{A}^1 - \lambda_{(I)} \boldsymbol{A}^0] \boldsymbol{r}_{(I)} = 0\,,
\end{equation}
has only real eigenvalues, $\lambda_{(I)}$, and $N$ independent, real
right-eigenvectors, $\boldsymbol{r}_{(I)}$ [see, \eg
  \cite{Anile_book}]. We will denote with $\boldsymbol{R}$ the matrix
of right eigenvectors, \ie
\begin{equation}
  R^I_{\phantom{I}J}  = r^I_{(J)}\,,
\end{equation}
and with $\boldsymbol{L}$ its inverse. We define the local
characteristic variables
\begin{align}
  &\boldsymbol{w} = \boldsymbol{L}\, \bu\,, &&\boldsymbol{Q} =
  \boldsymbol{L}\,\bF^1 \,,
\end{align}
and compute $\boldsymbol{Q}_{i+1/2,j,k}$ doing a component-wise
reconstruction, where $\boldsymbol{w}$ is used in place of $u$ and
$\boldsymbol{Q}$ in place of $f$ in the
(\ref{eq:lax.friedrichs.split}). Finally we set
\begin{equation}
  \bF^1_{i+1/2,j,k}  = \boldsymbol{R}\, \boldsymbol{Q}_{i+1/2,j,k}\,.
\end{equation}
This procedure is repeated in the other directions and yields the
wanted approximations of the $\partial \bF^a / \partial x^a$ terms in
$\bx_{i,j,k}$. All the results that we will be presented have been
obtained performing the reconstruction of the local characteristic
variables.

\subsection{Newtonian Hydrodynamics}

The equations of classical (\ie Newtonian) hydrodynamics describe the
conservation of mass, momentum and energy for a perfect fluid. They
can be written in the form (\ref{eq:claw}) with primitive variables,
\begin{equation}
  \bu = [\rho,\ \boldsymbol{v},\ \epsilon]\,,
\end{equation}
where $\rho$ is the density, $v^i$ the velocity and $\epsilon$ the
specific internal energy. The conserved variables are
\begin{equation}
\bF^{0}(\bu) = 
\left[\rho,\ \rho \boldsymbol{v},\ E \right] = 
\left[\rho,\ \rho \boldsymbol{v},\ 
\rho \big(\tfrac{1}{2}\, v^2 + \epsilon\big)\right]\,,
\end{equation}
the sources are zero and the fluxes are
\begin{equation}
  \bF^{i}(\bu) = \left[ \rho v^i,\ \rho v^i \boldsymbol{v} +
    p \pmb{\delta}^{i},\ v^i (E + p) \right],
\end{equation}
where $p$ is the pressure and $[\pmb{\delta}^i]^j = \delta^{ij}$ is
the Kronecker symbol. The system of equations is then closed by an
equation of state $p = p(\rho,\epsilon)$ and we adopt that of an ideal
fluid (or Gamma law)
\begin{equation}
\label{eq:ideal.gas.eos}
  p = (\Gamma-1)\rho\epsilon,
\end{equation}
where $\Gamma$ is the adiabatic index of the fluid. 

The Jacobians and their spectral decomposition for the equations of
Newtonian hydrodynamics and for a generic equation of state, can be
found, for example, in \cite{Kulikovskii2001}.

\subsection{Special-relativistic hydrodynamics}

In the case of the relativistic-hydrodynamic equations it is
convenient to work using a system of units in which $c = 1$ and we
adopt the standard convention for the summation over repeated
indices. We consider a perfect fluid having 4-velocity $\boldsymbol{u}
= (W ,\ W \boldsymbol{v})$, with $W \equiv (1-v^iv_i)^{-1/2}$ being
the Lorentz factor. Then the rest-mass current 4-vector is given by
\begin{equation}
  \boldsymbol{J} = \rho \boldsymbol{u},
\end{equation}
where $\rho$ is here the rest-mass density. The stress-energy tensor
is given by
\begin{equation}
  \boldsymbol{T} = 
\rho h \boldsymbol{u}\otimes\boldsymbol{u} + p\boldsymbol{g}\,,
\end{equation}
where $h = 1 + \epsilon + p / \rho$ is the specific enthalpy and
$\boldsymbol{g}$ is the spacetime metric, which we take to be that of
flat spacetime, \ie where the only nonzero components are the diagonal
ones and given by $g_{\mu\nu} = (-1,1,1,1)$. Because our main interest
with \texttt{THC} is of determining the statistical properties of
special-relativistic turbulence and of unveiling novel and
non-classical features (Radice \& Rezzolla in prep.) we will consider the fluid
not to affect the spacetime geometry, which we will consider to be
that of a flat spacetime at all times.

Conservation of rest-mass, momentum and energy are expressed by the
vanishing of the 4-divergence of $\boldsymbol{J}$ and $\boldsymbol{T}$
\begin{equation}
\label{eq:cons_eqs}
  \boldsymbol{\nabla} \cdot \boldsymbol{J} = 0\,, \qquad
  \boldsymbol{\nabla} \cdot \boldsymbol{T} = 0\,,
\end{equation}
where $\boldsymbol{\nabla}$ is the covariant derivative associated
with $\boldsymbol{g}$. Also in this case, the
relativistic-hydrodynamic equations~\eqref{eq:cons_eqs} can be cast in
the form (\ref{eq:claw}) with primitive variables
\begin{equation}
  \bu = [\rho,\ \boldsymbol{v},\ \epsilon]\,.
\end{equation}
The conservative variables are
\begin{equation}
  \bF^0(\bu) = [D,\ \boldsymbol{s},\ \tau]\,,
\end{equation}
where
\begin{equation}
  D \equiv \rho W\,, 
  \quad 
  \boldsymbol{s} \equiv \rho h W^2 \boldsymbol{v}\,, 
  \quad
  \tau \equiv \rho h W^2 - p - \rho W\,,
\end{equation}
and the fluxes are given by
\begin{equation}
  \bF^i(\bu) = 
[D v^i,\ \boldsymbol{s} v^i + p \pmb{\delta}^{i},\ s^i - D v^i ]\,.
\end{equation}

Finally, because we are working in special relativity, sources in the
system~\eqref{eq:cons_eqs} are zero and the equations are closed by an
equation of state, which we take again to be the ideal-fluid equation
of state~\eqref{eq:ideal.gas.eos}.

\begin{figure*}
  \begin{center}
    \includegraphics[width=17cm]{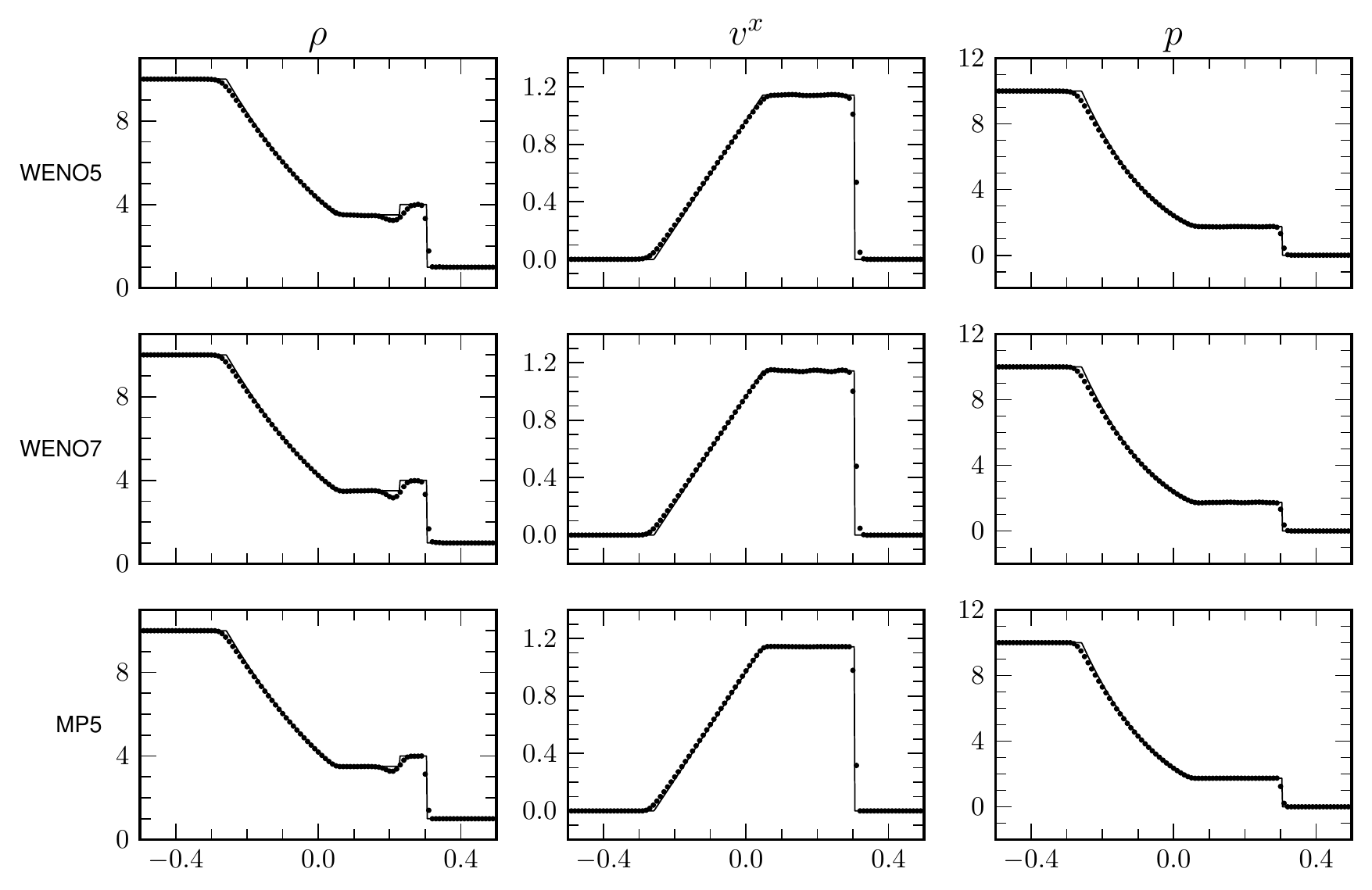}
  \end{center}
  \caption{Density (left panels) velocity (middle panels) and pressure
    (right panels) for the Newtonian strong shock test. We show both
    the analytic solution (solid line) and the numerical solutions
    obtained with different numerical schemes (dots). The resolution
    is $\Delta^1 = 1/100$ and the timestep is $\Delta^0 = 0.002$ for
    all the runs. Note that the numerical solution is not down-sampled
    and corresponds to a rather coarse resolution.}
  \label{fig:newtonian.strong.shock}
\end{figure*}

An important difference between the Newtonian and the
(special-)relativistic hydrodynamic equations is that in the latter
case there is no simple analytic expression for the inverse
transformation $\bF^0 \rightarrow \bu$, leading to the primitive
variables from the conserved ones. For this reason, we use a numerical
root-finding procedure to recover the primitive variables from the
conservatives. In particular we follow the strategy of
\cite{kastaun_2006_hrs,Kastaun2007}. The primitive variables can be easily
written as function of the conservative variables once a value for the
enthalpy, $\tilde{h}$, is assumed. At the same time the enthalpy can
be expressed as a function of the rest-mass density and of the
internal energy using the equation of state. Thus we can construct the
function
\begin{equation}
  g(\tilde{h}) = h_{\mathrm{EOS}}\big(\rho(\tilde{h}), \epsilon(\tilde{h})\big)
  - \tilde{h}\,,
\end{equation}
and use a one-dimensional root-finding procedure to find a
self-consistent value of the enthalpy, since $g(\tilde{h}) = 0$ if and
only if $\tilde{h}$ is the physical enthalpy. In particular, the
root-finding algorithm uses a combination of the Newton-Raphson method
with the regula-falsi and the bisection schemes: we check the
Newton-Raphson method for convergence and, in case of failure, we
switch to the regula-falsi. The bisection scheme is used as a
``fail-safe'' root-finder in situations where the regula-falsi is
converging too slowly: this is necessary only when the values of the
conservative variables are close to an unphysical region. In the large
majority of cases, the Newton-Raphson method usually converges to the
required level of accuracy with only three iterations on average.

The Jacobians and their spectral decomposition for the equations of
special-relativistic hydrodynamics and for a generic equation of
state, can be found in \cite{Donat98}.

\section{Numerical tests}\label{sec:tests}

This Section is dedicated to the presentation of some of the results
obtained with \texttt{THC} in a series of tests in Newtonian and
special-relativistic hydrodynamics.

\subsection{Newtonian hydrodynamics}

We begin with a series of tests in classical hydrodynamics, before
switching to the special-relativistic case.

\subsubsection{Strong shock}

The first test is a classical one-dimensional shock tube: the initial
data describes two regions filled with a $\Gamma=5/3$ ideal-fluid in
equilibrium separated by a membrane. At $t = 0$ the membrane is
removed and the two regions start to interact. The initial left and
right states are
\begin{align}
&   (\rho_L, v_L, p_L) = (10, 0, 10)\,, \quad
&&  (\rho_R, v_R, p_R) = ( 1, 0, 10^{-5})\,,
\end{align}
for $t > 0$ the analytic solution consists in a right-going shock
wave, followed by a right-going contact discontinuity and a transonic
left-going rarefaction wave.

In Figure \ref{fig:newtonian.strong.shock} we show with a solid line
the analytic solution, while filled circles are used to represent the
solution obtained at time $t =0.2$ with the different numerical
schemes. The grid resolution is $\Delta^1 = 1/100$ and the timestep
$\Delta^0 = 0.002$ for all the runs. Note that the numerical solution
is not down-sampled and hence it corresponds to a genuinely coarse
resolution. We do not show the results obtained with the
bandwidth-optimised schemes since they are basically indistinguishable
from the ones obtained using their traditional counterpart, \ie the
solution obtained with WENO3B is basically on top of the one obtained
with WENO5 and the solution obtained with WENO4B is identical, at the
plot scale, with the one obtained with WENO7. Since the results
obtained with the bandwidth-optimized schemes are found to be very
close to the ones obtained with the standard WENO schemes in all the
shock-tube test that we performed, we will show only the numerical
solutions obtained with WENO5, WENO7 and MP5 in all the cases.

\begin{figure*}
  \begin{center}
    \includegraphics[width=17cm]{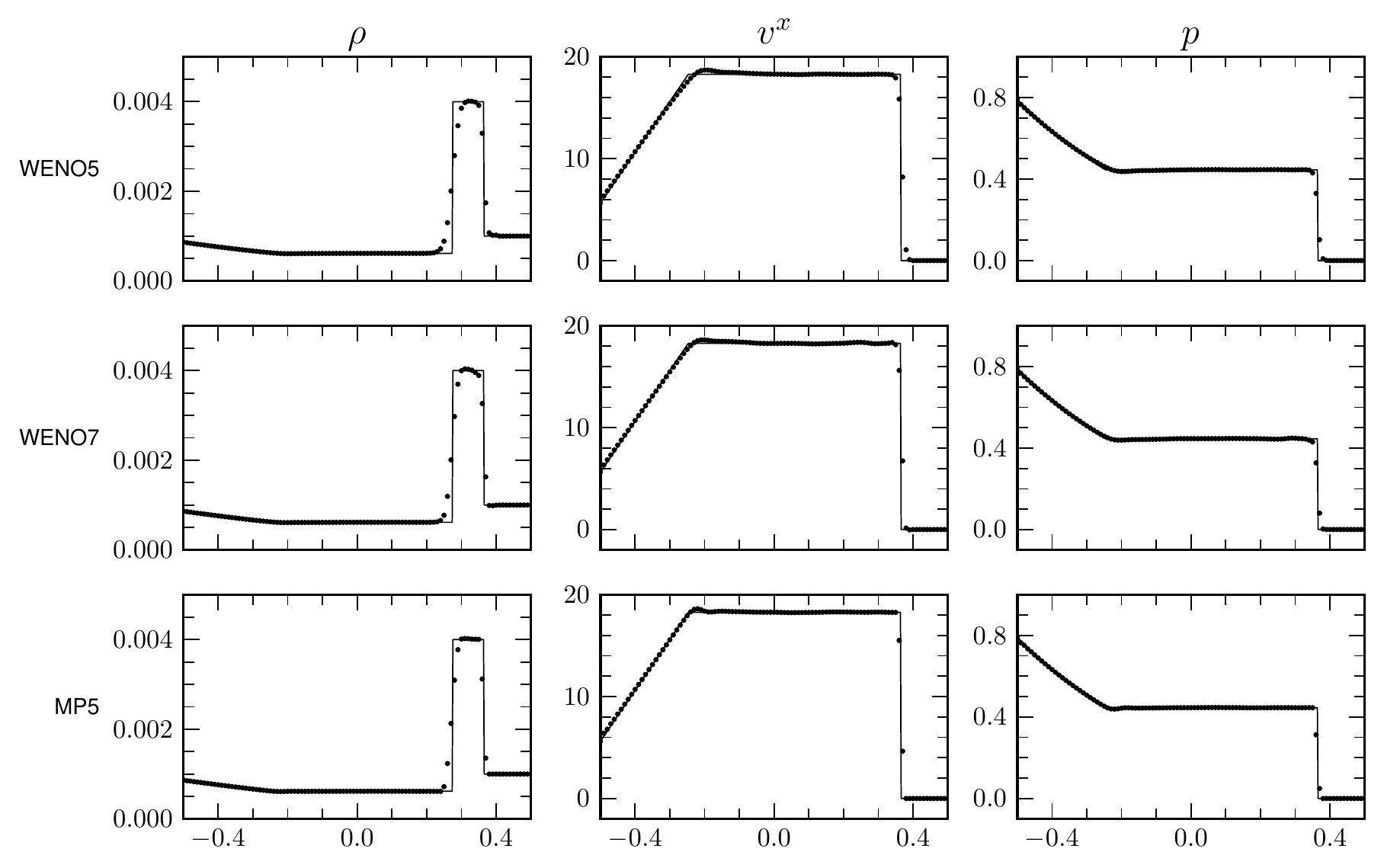}
  \end{center}
  \caption{The same as in Figure \ref{fig:newtonian.strong.shock}, but for the
  Newtonian blast-wave test. The resolution is $\Delta^1 = 1/100$ and
  the timestep is $\Delta^0 = 0.0001$ for all the runs.}
  \label{fig:newtonian.blastwave}
\end{figure*}

Even at this fairly low resolution, all the schemes are able to capture
well both the shock wave and the rarefaction wave, showing the good
behaviour of our entropy fix. The shock wave is captured within three
gridpoints. There are no appreciable post-shock oscillations in the
solution obtained with the MP5 reconstruction, while small
oscillations are present in the velocity field with WENO5 and, in
particular, with WENO7.

We note that the contact discontinuity is resolved, but not without
oscillations. However, we should bear in mind that, although the
density contrast is small across the contact discontinuity, this test
is a severe one due to the high Mach number of the shock wave, \ie
$\MM_s \approx 360$.

\subsubsection{Blast wave}
\label{sec:newtonian.blastwave}

The second test is similar to the first one, but results in a much
larger density contrast at the contact discontinuity. The initial data
is given by
\begin{align}
&  (\rho_L, v_L, p_L) = (10^{-3}, 0, 1)\,, \quad
&& (\rho_R, v_R, p_R) = (10^{-3}, 0, 10^{-5})\,,
\end{align}
and the adiabatic index is still $\Gamma=5/3$. Also in this case, the
analytic solution consists in a right-going shock wave, followed by a
contact discontinuity and a left-going rarefaction wave.

In Figure \ref{fig:newtonian.blastwave} we show the exact solution at
time $t=0.015$ (solid line), as well as the numerical solution (filled
circles) obtained using different numerical schemes. The grid
resolution is $\Delta^1 = 0.01$ ($N = 100$) and the timestep is
$\Delta^0 = 0.0001$ for all the runs. The Mach number of the shock
wave is $\MM_s \approx 190$ and in this case all our schemes are free
from major oscillations. The shock wave is resolved within two grid
points, while the contact discontinuity is smeared over 5-6 grid
points.

The MP5 scheme is able to properly capture the constant state between
the shock wave and the contact discontinuity, while the WENO schemes
result in more ``rounded'' solutions. In particular both WENO5 and
WENO7 overestimated the density contrast.

\begin{figure*}
  \begin{center}
    \includegraphics[width=17cm]{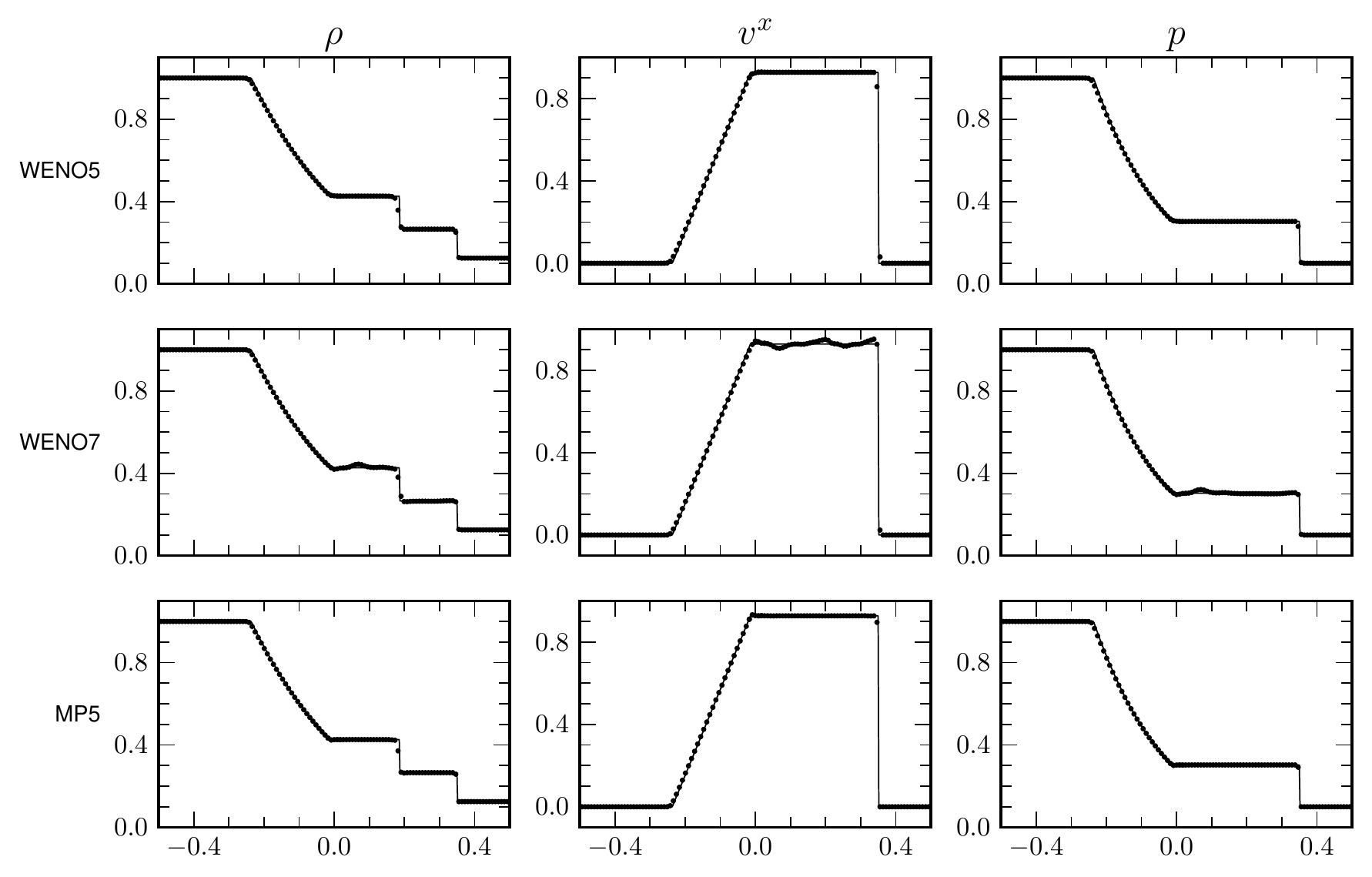}
  \end{center}
  \caption{The same as in Figure \ref{fig:newtonian.strong.shock}, but for the
  Newtonian rotated Sod test. The resolution is $\Delta ^i = 1/200$ and the
  timestep is $\Delta^0 = 0.000625$ for all the runs.}
  \label{fig:newtonian.diagonal.sod}
\end{figure*}

\subsubsection{Rotated Sod test}

A genuinely three-dimensional shock-tube test in Newtonian
hydrodynamics is offered by the classical rotated Sod test
\citep{Sod1978}. In this case, the adiabatic index is $\Gamma=1.4$ and
the right and left states are
\begin{align}
&  (\rho_L, v_L, p_L) = (1, 0, 1)\,, \quad
&& (\rho_R, v_R, p_R) = (0.125, 0, 0.1)\,,
\end{align}
the initial, 1D data, is rotated by $45^\circ$ about the $z$ and $y$
axes to yield a shock wave that is diagonal to the principal axes of
the grid. The analytic solution consists in a left-going rarefaction
wave and a right-going shock wave separated by a right-going contact
discontinuity.

In Figure \ref{fig:newtonian.diagonal.sod} we show the analytic
solution in the diagonal direction (solid line), as well as the
numerical solutions (filled circles), at time $t = 0.2$. The spatial
resolution is $\Delta^i = 1/200$ for $i = 1,2,3$ and the timestep is
$\Delta^0 = 0.000625$ for all the runs. All the schemes are able to
properly capture the main features of the solution: the
discontinuities are captured within 1 or 2 gridpoints and both WENO5
and MP5 are able to capture the plateau in the velocity. The solution
obtained with the WENO7 scheme presents oscillations in the velocity
after the shock wave and in the density and pressure fields downstream
the contact discontinuity. Similar oscillations are also observed with
WENO5 and MP5 when the resolution is halved.

Overall, these tests demonstrate the accuracy of the dimensionally
unsplit approach that we use to treat the multi-dimensional case.

\begin{figure*}
  \begin{minipage}{0.49\hsize}
    \includegraphics[width=\textwidth]{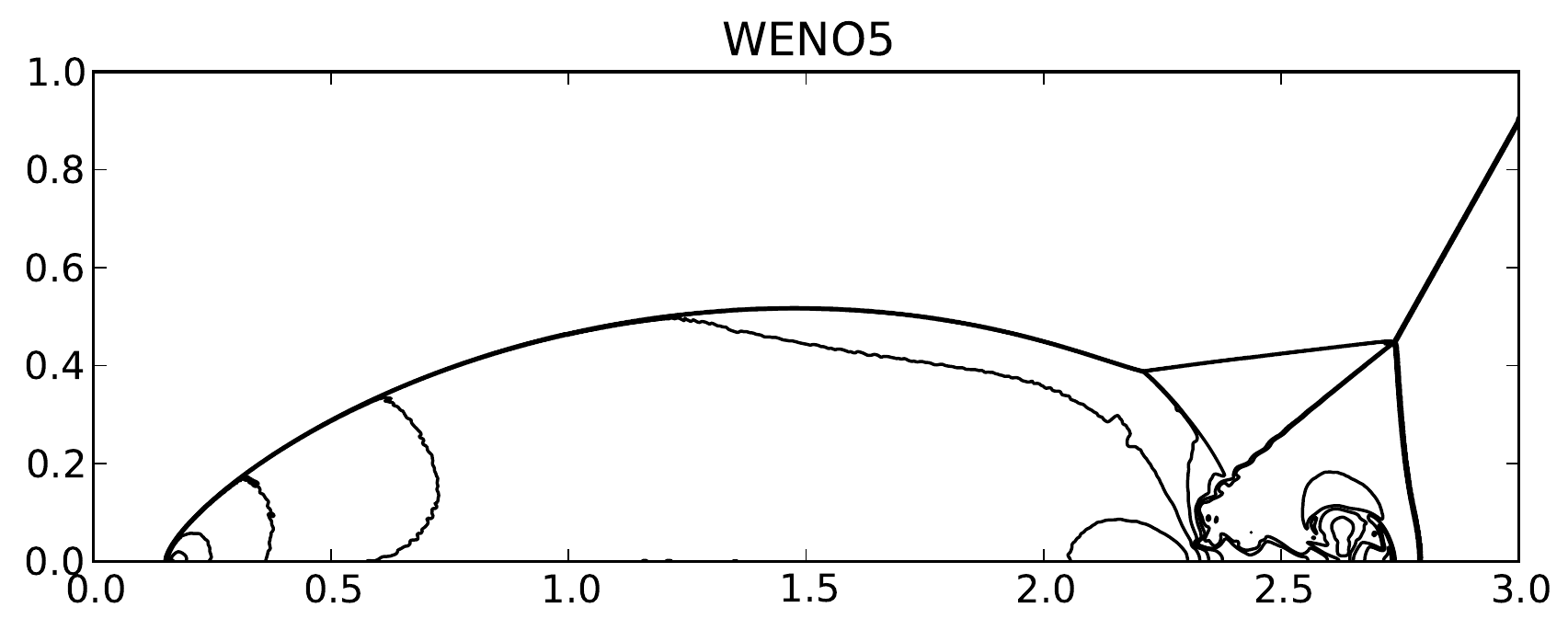}
    \includegraphics[width=\textwidth]{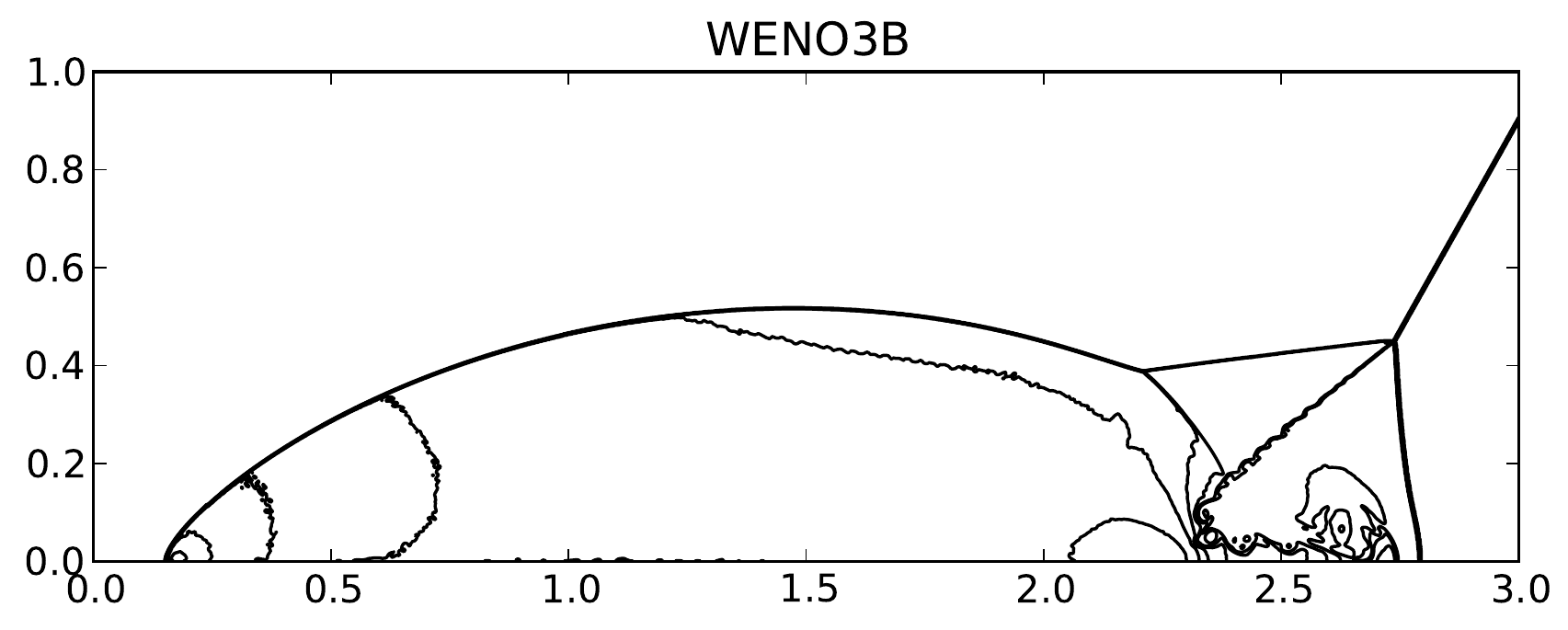}
  \end{minipage}
  \begin{minipage}{0.49\hsize}
    \includegraphics[width=\textwidth]{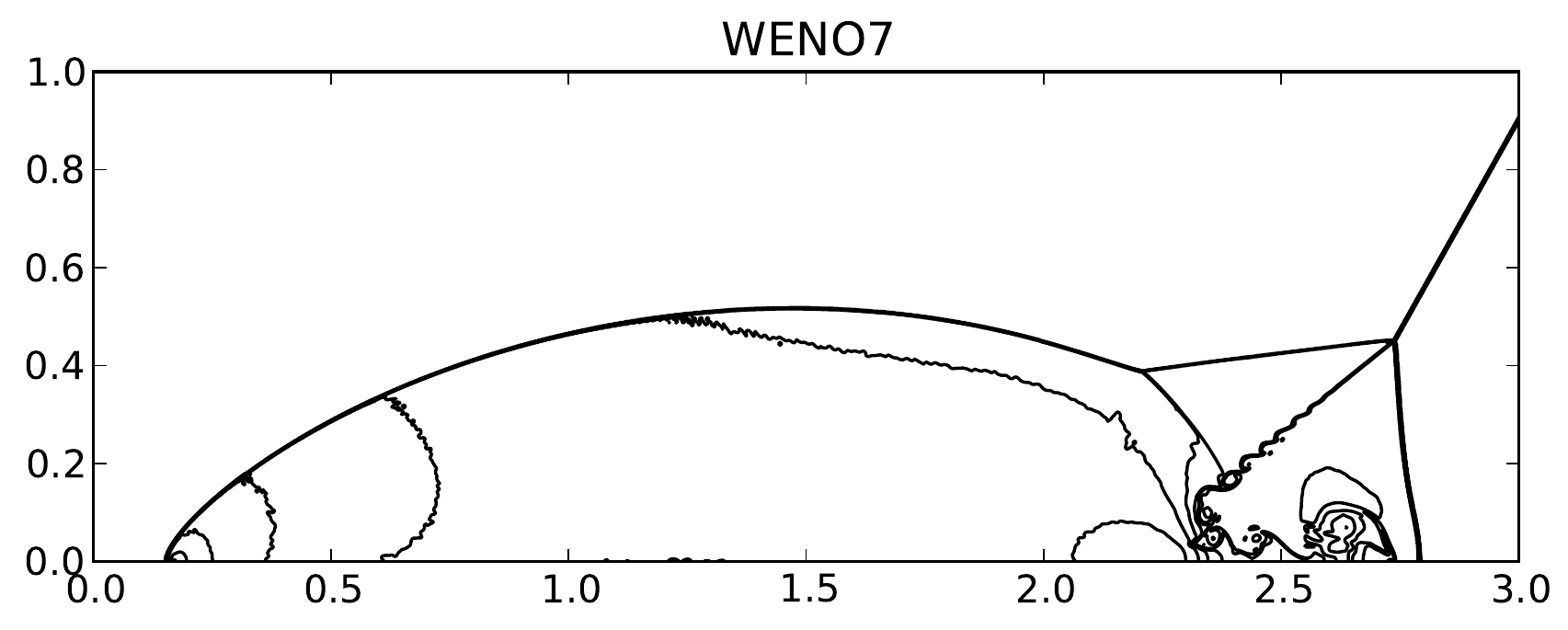}
    \includegraphics[width=\textwidth]{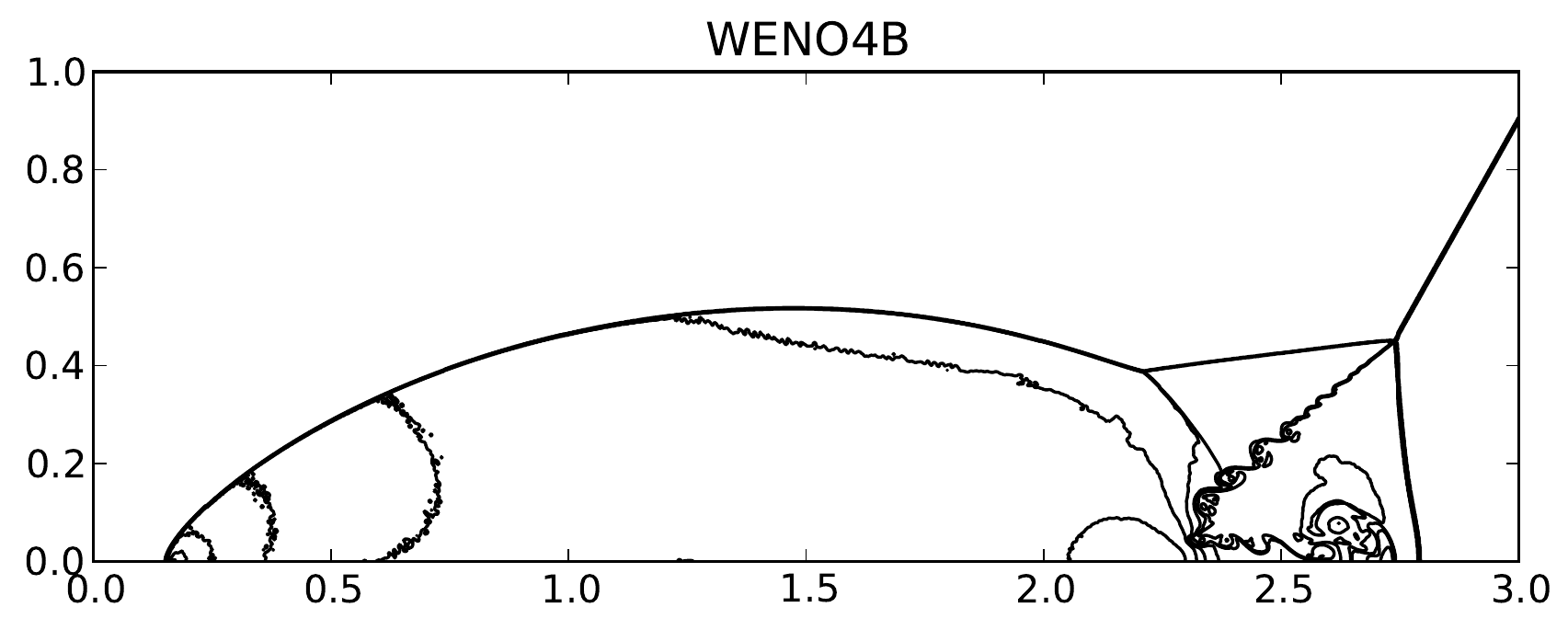}
  \end{minipage}
  \begin{center}
    \includegraphics[width=0.49\textwidth]{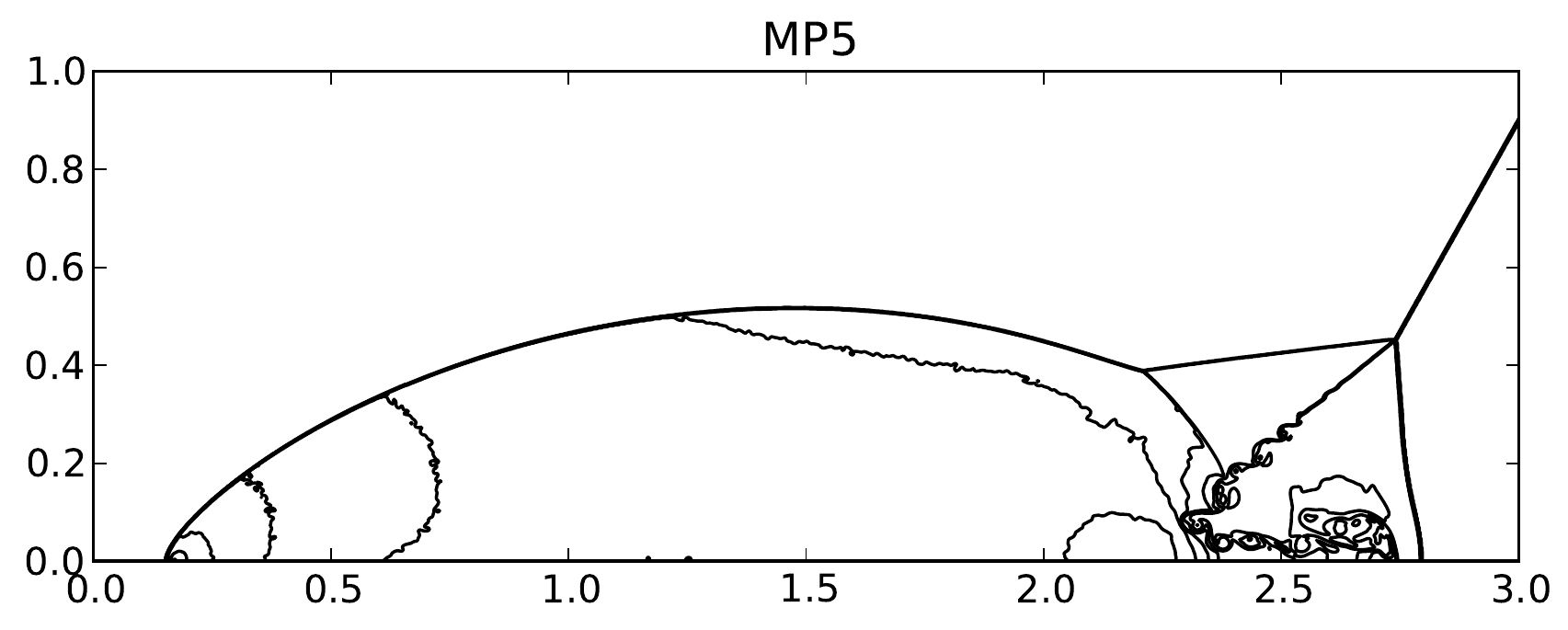}
  \end{center}
  \caption{Isocontours of the rest-mass density at time $t=0.2$ for
    the Newtonian double Mach reflection test, obtained with different
    schemes. We show 20 contours levels equally spaced between 0 and
    3. The resolution is $\Delta^i = 1 / 480$ and the timestep is
    $\Delta^0 = 1 / 40000$ for all the runs.}
  \label{fig:newtonian.double.mach}
\end{figure*}

\subsubsection{Double Mach reflection test}

A final test in Newtonian hydrodynamics is the double-Mach reflection
test proposed by \cite{Woodward84}. The initial data describes a
right-going Mach 10 shock wave making a $60^\circ$ angle with the
computational grid and intersecting the $x-$axis at $x = 1/6$. A
perfectly reflecting wall is placed along the $x-$axis in the $x >
1/6$ region, while the values along the other regions of the boundary
are set to the pre and post-shocked values on the left/right side of
the shock wave [see \cite{Woodward84} for details on the boundary
  conditions and the initial data].

We considered the numerical solutions obtained with WENO5, WENO7, MP5,
WENO3B and WENO4B in the computational domain $0 \leq x \leq 4$, $0
\leq y \leq 1$. For each of these schemes we performed five runs with
resolutions $120 \times 30$, $240 \times 60$, $480 \times 120$, $960
\times 240$ and $1920 \times 480$. In Figure
\ref{fig:newtonian.double.mach} we show isocontours (with an equal
spacing of 0.15 between 0 and 3) of the rest-mass density obtained at
the highest resolution at time $t = 0.2$ by the different numerical
schemes. In order to ease the comparison with the results reported in
\cite{Woodward84}, we show only the region $x < 3$.

Our code performs well in this test across all the reconstruction
schemes that we tried. All the discontinuities, including the contact
discontinuity ahead of the jet region, are well resolved within few
grid regions, even at the lowest resolution. As we increase the number
of gridpoints we do not observe any sign of the carbuncle phenomenon
and this seems to be an indication that our algorithm is able to
introduce enough numerical dissipation to avoid the odd-even
decoupling.

At high-enough resolution it is possible to observe the development of
instabilities upstream from the reflected shock wave, generating small
scale structures along the contact discontinuity. The ability of the
different codes to resolve these structures can be used as an
indication of their numerical viscosity. In particular we can see how
the bandwidth-optimized schemes gain with respect to their
``standard'' counterparts. For instance, WENO3B, which uses the same
stencil as WENO5, yields a solution which is in qualitative agreement
with the one obtained using WENO5 at twice the spatial resolution. The
same is also true if we compare WENO4B and WENO7.

All things considered, we find that the best performance is given by
the MP5 scheme. This algorithm has a computational cost which is
comparable with the one of WENO5, as it uses the same
stencil. Nevertheless the solution obtained with MP5 is very close to
the one obtained with the optimized WENO4B scheme, which, in 3D, is
almost twice as expensive as WENO5.

%
%

\subsection{Special-relativistic hydrodynamics}

In this Section we will present the results obtained in a series of
tests in special-relativistic hydrodynamics.

\begin{figure}
  \resizebox{\hsize}{!}{
  \includegraphics{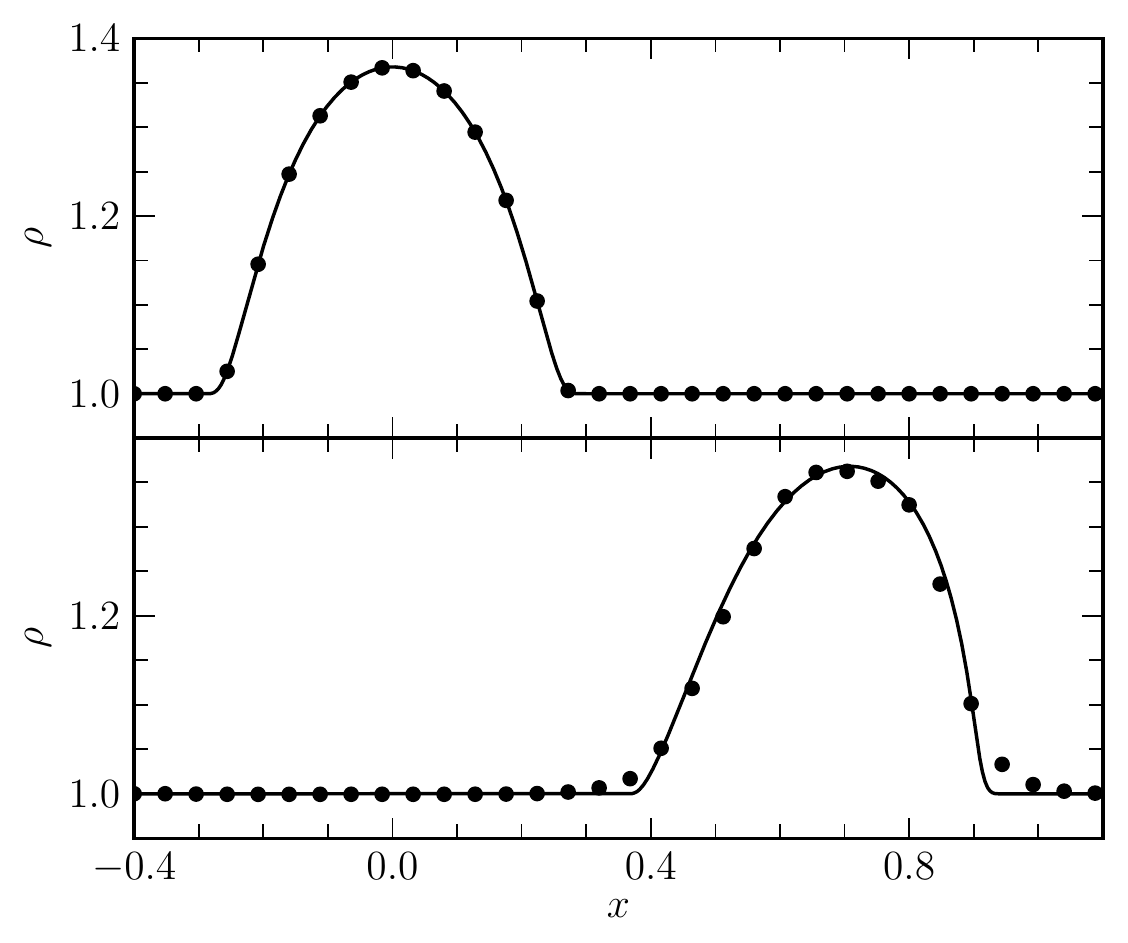}}
  \caption{Analytic solution and numerical solution computed with
    WENO3B with $50$ gridpoints for the case of an smooth wave in an
    adiabatic relativistic fluid. The figure shows both the initial
    data (top panel) and the solution at time $t=0.8$ (bottom
    panel). The solid line represent the analytic solution, while the
    filled circles represent the numerical one.}
  \label{fig:relativistic.smooth.wave.rho}
\end{figure}

\subsubsection{Adiabatic smooth flow}

The first test we present is designed to show the accuracy of the code
in the case of smooth solutions and hence to measure rigorously the
convergence order of \texttt{THC} for the different schemes
implemented. This test is very similar to the one discussed by
\cite{Zhang2006}.

We consider a one-dimensional, large-amplitude, smooth, wave
propagating in an isentropic fluid, with polytropic equation of state,
\begin{equation}
  p = K \rho^\Gamma,
\end{equation}
where $K = 100$ and $\Gamma = 5/3$. The rest-mass density at $t = 0$ is
given by
\begin{equation}
  \rho_0(x) =
  \begin{cases}
  1 + \exp\left[-1/(1 - x^2/L^2)\right]\,, & \textrm{if } |x| < L; \\
  1, & \textrm{elsewhere};
  \end{cases}
\end{equation}
where, differently from \cite{Zhang2006}, the initial profile of the
rest-mass density is chosen to be $C^\infty$ but is not analytic. We
have found this choice important to obtain the correct convergence
order at very high resolutions. Indeed, when adopting the same profile
as in \cite{Zhang2006}, we found that the jump discontinuity in the
fifth derivative of the initial data prevents the WENO7 scheme from
achieving a convergence order larger then five. Besides this small
difference, our initial data is basically identical to the one used by
\cite{Zhang2006}. The initial velocity is setup assuming that one
of the two Riemann invariants~\citep{Anile_book},
\begin{equation}
  J_- = \frac{1}{2} \ln\left(\frac{1+v}{1-v}\right) - \frac{1}{\sqrt{\Gamma-1}}
    \ln\left(\frac{\sqrt{\Gamma-1}+c_s}{\sqrt{\Gamma-1}-c_s}\right)\,,
\end{equation}
where $c_s$ is the sound speed, is constant in the whole region and so
that $v = 0$ if $|x| \geq L$. The other Riemann invariant,
\begin{equation}
  J_+ = \frac{1}{2} \ln\left(\frac{1+v}{1-v}\right) + \frac{1}{\sqrt{\Gamma-1}}
    \ln\left(\frac{\sqrt{\Gamma-1}+c_s}{\sqrt{\Gamma-1}-c_s}\right)\,,
\end{equation}
is not constant, so that the initial data describes a right-going wave.

The analytic solution can be easily found in Lagrangian coordinates,
up to the caustic point, using the method of characteristics
\citep{Anile_book}, while its calculation in Eulerian coordinates
involves the solution of a transcendental equation. For the purposes
of our calculation, a good-enough approximation of the exact solution
was obtained by computing it on a very fine Lagrangian grid (we have
used $100,000$ gridpoints), and interpolated on the Eulerian grid
using a cubic spline interpolation. This solution is then used as the
reference solution against which the numerical solutions obtained with
\texttt{THC} have been compared.

In our test we set $L = 0.3$ and use a computational grid in the
region $-0.4 \leq x \leq 2$, evolving the initial data up to time $t =
1.6$, which is approximately the time when a caustic appears and the
solution becomes discontinuous. We performed this test using different
schemes and different resolutions and we measured the $L^1-$norm of
the error against the reference solution. Differently from all the
other tests, instead of the third-order SSP-RK scheme, we adopt here a
fourth-order RK time integrator. The reason for this choice is that,
since this test involves a smooth solution, a SSP time-integrator is
not required. Moreover, as we will show in the following, the use of a
more accurate time integrator enabled us to measure a convergence
order of the spatial discretization which is not spoiled by the order
in time, the only exception being the highest resolution run done with
WENO7. The CFL factor is set to be $C \approx 0.2$.

The initial rest-mass density profile. as well as the solution at time
$t=0.8$, which we take as a reference time for the measure of the
error, is shown in Figure \ref{fig:relativistic.smooth.wave.rho},
together with the solution obtained with WENO3B using $50$
gridpoints. As it can be seen from the Figure, the solution at the
considered time is still smooth and our scheme is able to properly
capture it very well even at this very coarse resolution.

\begin{figure}
  \resizebox{\hsize}{!}{%
  \includegraphics{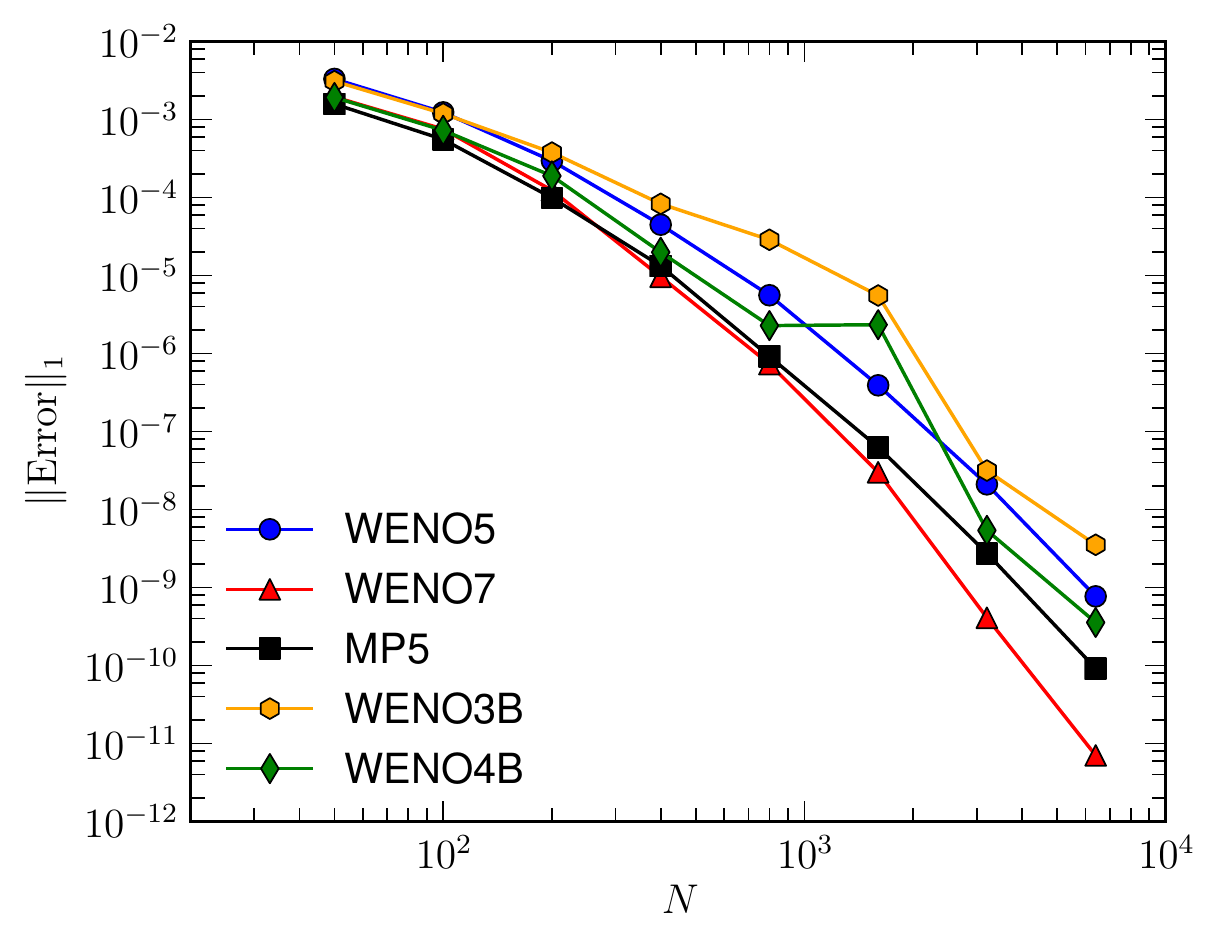}}
  \caption{$L^1-$norm of the error for different resolutions and for
  different numerical schemes for the case of a smooth simple wave in an
  adiabatic relativistic fluid}
  \label{fig.relativistic.smooth.wave.convergence}
\end{figure}
%
%

The $L^1-$norm of the error, as measured at the reference time, is
shown in Figure \ref{fig.relativistic.smooth.wave.convergence}. The
first thing one notices is that all our schemes approach the expected
convergence order only asymptotically, at very high resolution.
The reason for this behaviour is in the ``kinks'' ahead and behind the
pulse, where the numerical error is largest. These regions are
``misinterpreted'' as discontinuities by the shock-detection part of
our schemes, unless they are resolved with enough gridpoints.

The best performing scheme in this test is the MP5 one: at low
resolution it yields a smaller error then the seventh-order WENO
scheme, which, in turn, is able to attain an higher convergence order
only at a resolution which is unfeasible in any practical
multidimensional applications. The bandwidth optimized schemes present
a somewhat errant behaviour in their convergence order, with WENO4B,
not even showing a monotone trend in $L^1-$norm of the error as a
function of the number of gridpoint. We do not presently have an 
explanation we find sufficiently convincing for the behaviour shown.

\begin{figure}
  \resizebox{0.94\hsize}{!}{%
  \includegraphics{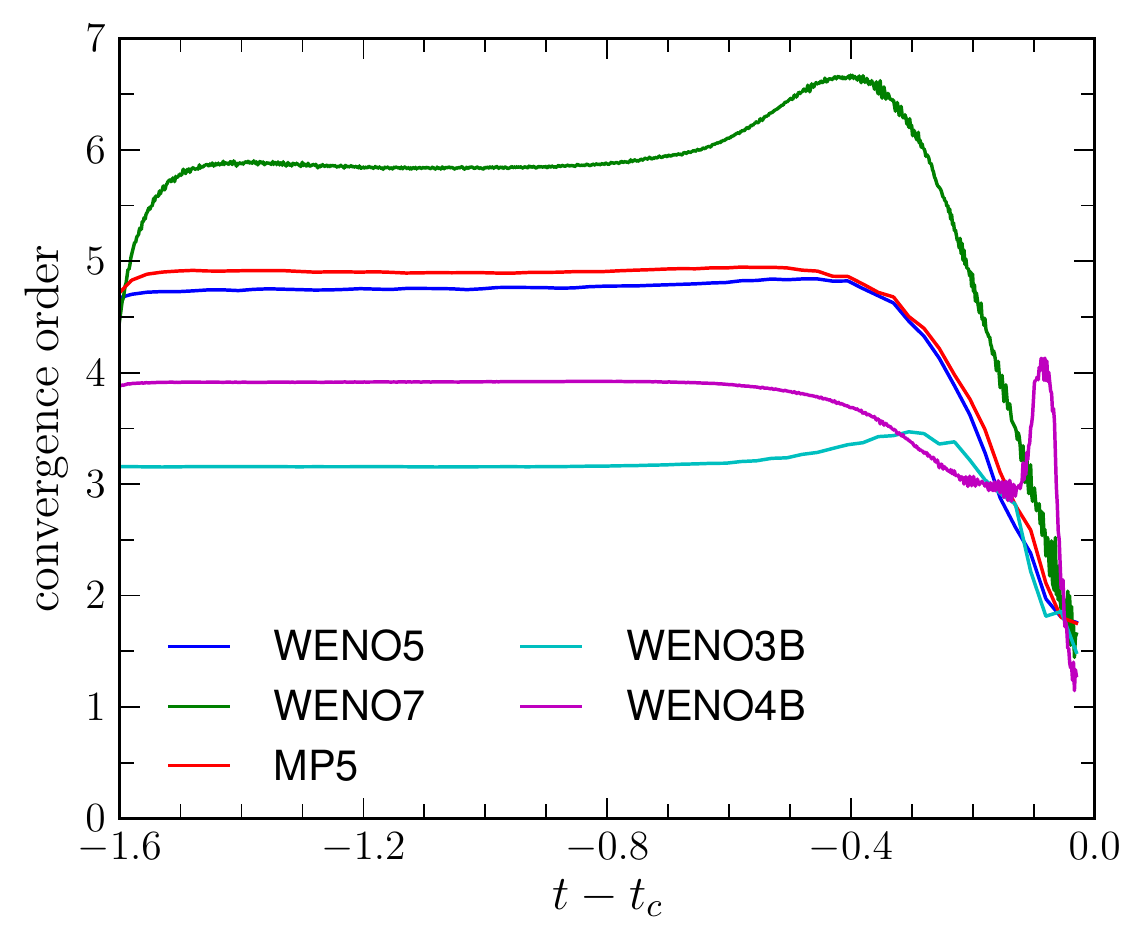}}
  \caption{Order of convergence, as measured using the two highest resolution
  runs, as a function of the time to the caustic for
  different numerical schemes for the smooth simple wave test.}
  \label{fig.relativistic.smooth.wave.order}
\end{figure}

Finally it is interesting to study the convergence order of the different
numerical schemes as a function of $t - t_c$, where $t_c$ is the time
when the caustic is formed. The convergence order is computed using the
error with respect to the exact solution of the two highest resolution
runs and is shown in Figure \ref{fig.relativistic.smooth.wave.order}. At
these very high resolutions, all the schemes appear to be converging at
their nominal converge order away from the caustic, apart from WENO7 that
appears to have already reached saturation. Its order of convergence
increases up to almost seven close to the caustic, at time $t-t_c \approx
-0.4$, because there the error is dominated by the presence of a more
steep gradient ahead of the pulse. Before that, the error from the
spatial discretization is probably very close to the one from the time
discretization (we recall that we use a fourth-order RK integrator), thus
degrading the convergence order.

Indeed, as the time of shock formation approaches, the order of the
schemes decreases slowly to the first-order expected in the case of
discontinuities. The bandwidth optimized schemes and, in particular,
WENO4B show, again, an erratic behaviour of their convergence
order. The reasons for this behaviour are most probably the same ones
behind the similar behaviour observed while looking at the error as a
function of the number of gridpoints.

%

%
\begin{figure*}
  \begin{center}
    \includegraphics[width=17cm]{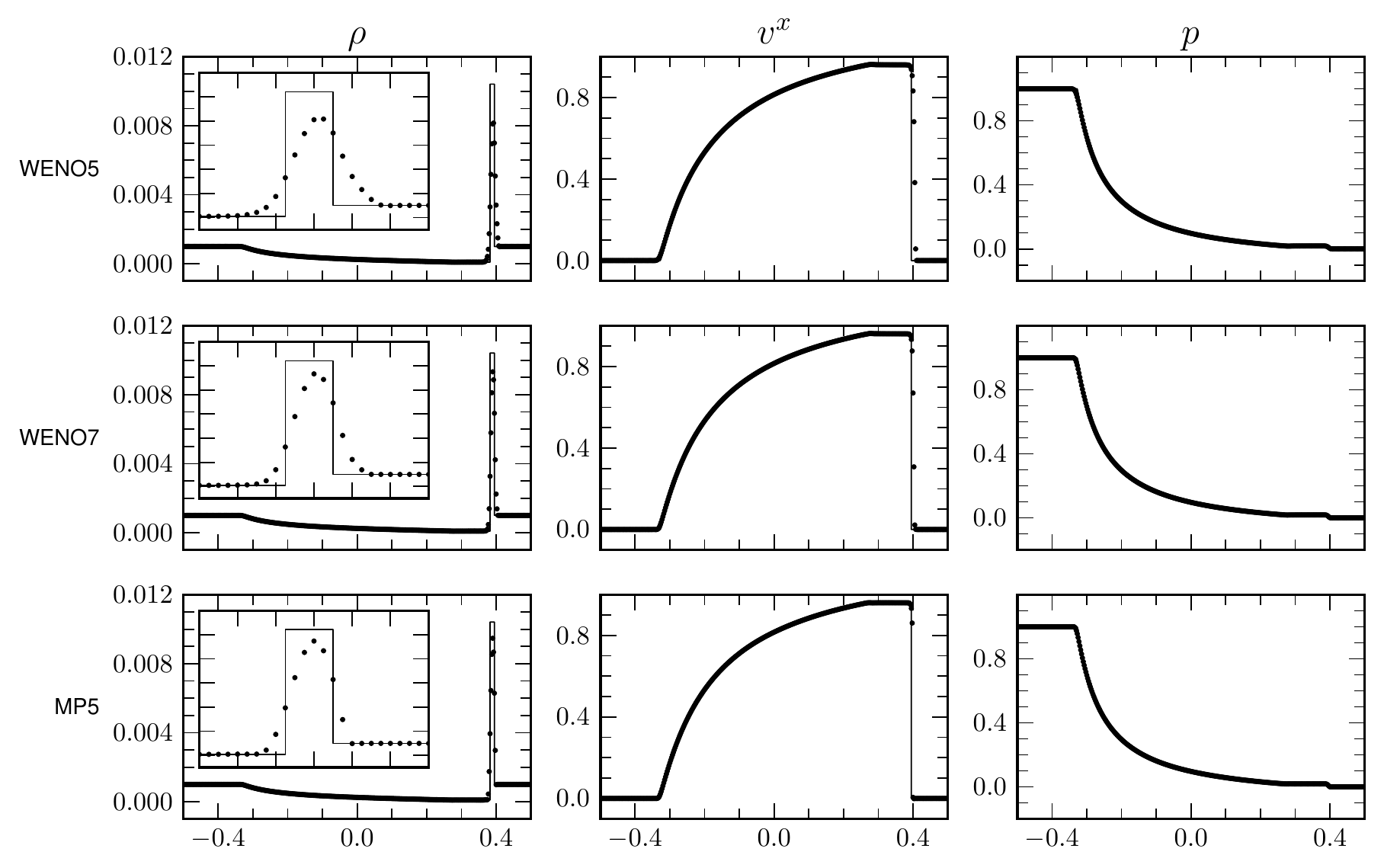}
  \end{center}
  \caption{The same as in Figure \ref{fig:newtonian.strong.shock}, but for
    the relativistic blast-wave test. The resolution is $\Delta^1 =
    1/400$ and the timestep is $\Delta^0 = 0.0005$ for all the runs.}
  \label{fig:relativistic.blastwave}
\end{figure*}

\subsubsection{Blast wave}
\label{sec:relativistic.blastwave}

When contrasted with its Newtonian counterpart, one of the most
striking features of relativistic hydrodynamics is that relativistic
fluids can exhibit much stronger shock waves. For this reason, it is
important to assess the capability of the code to handle very strong
shocks. As a first example we consider a one-dimensional shock-tube
where the initial data is given by,
\begin{align}
&  (\rho_L, v_L, p_L) = (10^{-3}, 0, 1)\,, \quad
&& (\rho_R, v_R, p_R) = (10^{-3}, 0, 10^{-5})\,,
\end{align}
and the adiabatic index is still $\Gamma=5/3$. This initial data is
formally identical to the one used in Section
\ref{sec:newtonian.blastwave} in the Newtonian case. The analytic
solution consists again in a transonic left-going rarefaction wave and
in a right-going shock wave separated by a right going contact
discontinuity. The shock wave has a relativistic Mach number
$\mathcal{M}_s \approx 50$ \citep{Chiu1973, Konigl1980}.

The results obtained with the different numerical schemes are reported
in Figure \ref{fig:relativistic.blastwave}, where we show the analytic
solution (solid line) as well as the numerical ones (filled circles)
obtained with WENO5, WENO7 and MP5, at time $t=0.4$. As in the
Newtonian case, we do not show the results obtained with the bandwidth
optimized schemes as they are basically identical to the ones of their
traditional variant. The spatial resolution that we use is $\Delta^1 =
1/400$ and the CFL factor is $C = 1/5$ in all cases.

\begin{figure*}
  \begin{center}
    \includegraphics[width=17cm]{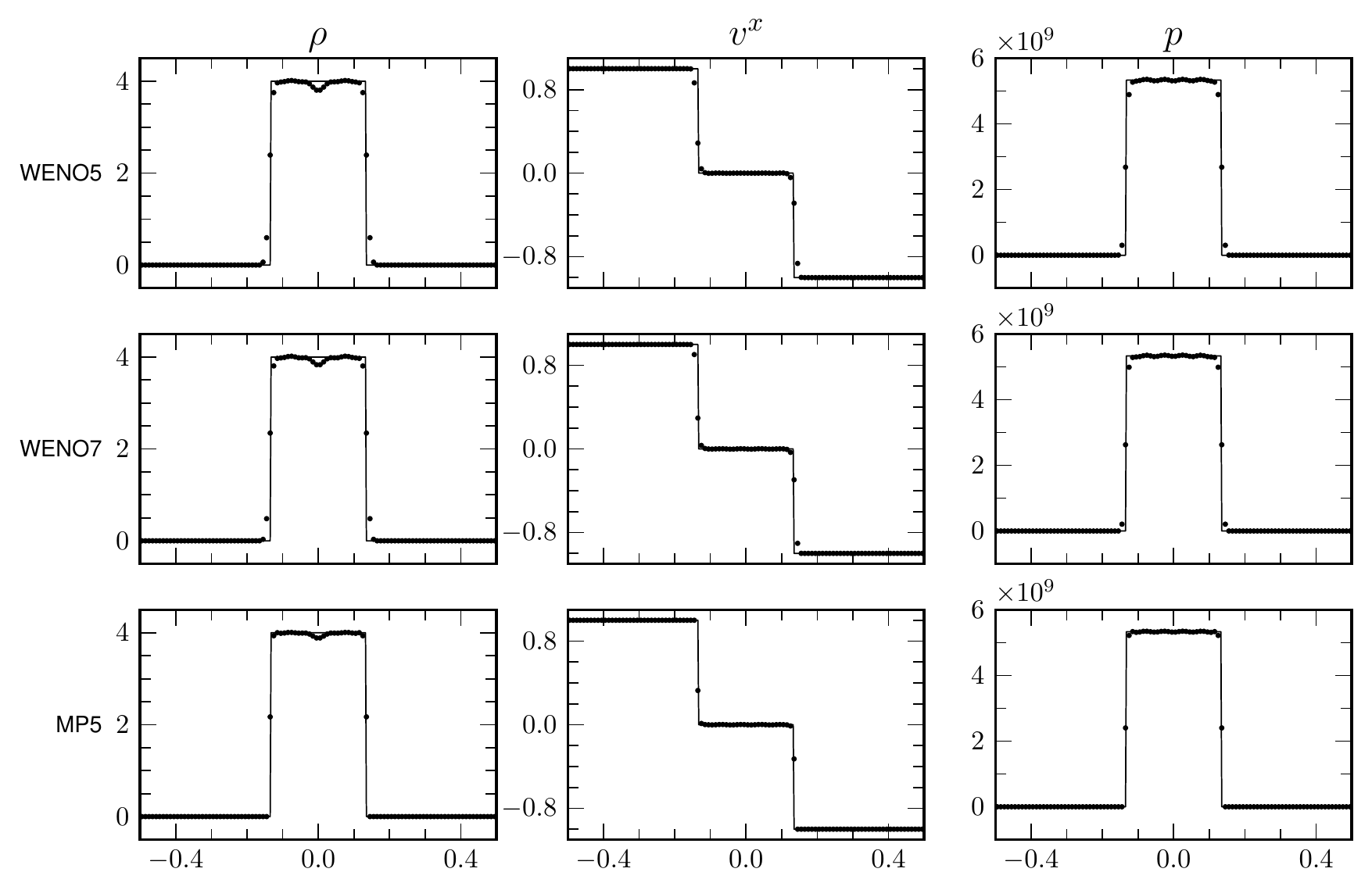}
  \end{center}
  \caption{The same as in Figure \ref{fig:newtonian.strong.shock}, but
    for the relativistic shock-heating test. The resolution is
    $\Delta^1 = 1/100$ and the timestep is $\Delta^0 = 0.001$ for all
    the runs.}
  \label{fig:relativistic.implosion}
\end{figure*}

The CFL factor in this test is basically constrained by the MP5
scheme: while the WENO schemes appear to be robust and stable up to
CFL factor of $C \approx 2/5$, the MP5 algorithm produces large
oscillations and yields non-physical values in the conservative
variables unless a smaller CFL factor is used. We point out that for
this particular test the necessity of using small timesteps to avoid
the creation of unphysical states has been observed also when using
other numerical schemes. For instance, when using the finite-volume
code \texttt{Whisky} \citep{Baiotti03a,Baiotti04}, with the HLLE
approximate Riemann solver [see, \eg \cite{Toro99}] and the PPM
reconstruction \citep{colella_1984_ppm}, the maximum allowed CFL
factor was also found to be $C \approx 2/5$. It is also known that,
for the monotonicity-preserving property to hold for the MP5 scheme,
the timestep must be subject to an additional constraint which is
distinct from the standard CFL condition \citep{suresh_1997_amp}. Yet,
it is somewhat surprising to observe that MP5 requires a timestep
which is smaller by a factor of order two with respect to the WENO
schemes. Furthermore, this property does not seem to be a peculiarity
of this specific problem, since we observed a similar behaviour also
for the other tests that we performed.

\cite{Zhang2006} report that the use of MP5 reconstruction results in
large numerical oscillations in their special relativistic code. In
our code, we see a similar behaviour unless we use a timestep which is
about half of the one considered ``safe'' for the WENO scheme. If the
timestep is sufficiently small, on the other hand, the MP5 algorithm
results in very accurate solutions, as in the Newtonian case. In
particular in Figure \ref{fig:relativistic.blastwave} we can see the
results obtained with this particular test. Our code, with MP5, is
able to capture the shock wave within three gridpoints while the
contact discontinuity is spread across six points. WENO7 yields a
solution of similar quality, while WENO5 is slightly more diffusive.

The capability of a numerical code to capture the density contrast for
this particular test is a classical benchmark for relativistic
hydrodynamics codes. Again, using as reference the \texttt{Whisky}
code, the use of the HLLE solver with PPM reconstruction and with
artificial compression, leads to a maximum density which is $71\%$ of
the analytic solution. At the same resolution, our \texttt{THC} using
a WENO5, WENO7 and MP5 reconstruction is able to attain a maximum
density which is respectively $78\%$, $90\%$ and $91\%$ of the
analytical value. These results are in very good agreement with the
ones reported by \cite{Zhang2006}, who measured a relative value of
$72\%$ and $79\%$ for their implementation of PPM and WENO5 schemes.

\subsubsection{Shock-heating}

An even more striking example of how relativistic effect can enhance
the density contrasts in shock waves is given by the classical
shock-heating test. In this case, the initial data is given by
\begin{align}
&  (\rho_L, v_L, p_L) = (10^{-3}, v, 10^3)\,, \quad
&& (\rho_R, v_R, p_R) = (10^{-3}, -v, 10^3)\,,
\end{align}
the polytropic index is $\Gamma = 4/3$ and
\begin{equation}
  v = \sqrt{1-\frac{1}{W^2}} \simeq 0.99999949\,,
\end{equation}
where the Lorentz factor is set to be $W = 1000$. In this case, the
analytic solution is represented by two shocks whose collision
compresses the fluid, converting its kinetic energy into thermal
energy, that is, through a ``shock heating''.

In Newtonian hydrodynamics the maximum compression ratio can be
computed as
\begin{equation}
  \sigma_{\mathrm{newt}} = \frac{\Gamma + 1}{\Gamma - 1} = 7\,,
\end{equation}
for all the values of $v$, while it is easy to show \citep{Marti03}
that in the relativistic case the compression ratio is
\begin{equation}
  \sigma = \frac{\Gamma + 1}{\Gamma - 1} + 
\frac{\Gamma}{\Gamma - 1} (W - 1) \simeq 4003\,,
\end{equation}
thus about three orders of magnitude larger for the same adiabatic
index and growing linearly with the Lorentz factor.

The exact solution at time $t=0.4$, as well as the numerical solutions
obtained with our code, are shown in Figure
\ref{fig:relativistic.implosion}. As it can be seen from the Figure,
\texttt{THC} is able to handle very well this large compression
ratio. The WENO5 and WENO7 solutions are affected by some small
wall-heating effect \citep{noh_1987_ecs, rider_2000_rwh}, resulting in
a slight under-density near $x=0$ and some numerical oscillations in
the pressure. The MP5 scheme, on the other hand, yields a solution
which is essentially free from oscillations in the pressure and much
less affected by the wall-heating effect in the density variable,
although at the cost of a smaller timestep, as discussed in the
previous Section.

\begin{figure*}
  \begin{center}
    \includegraphics[width=17cm]{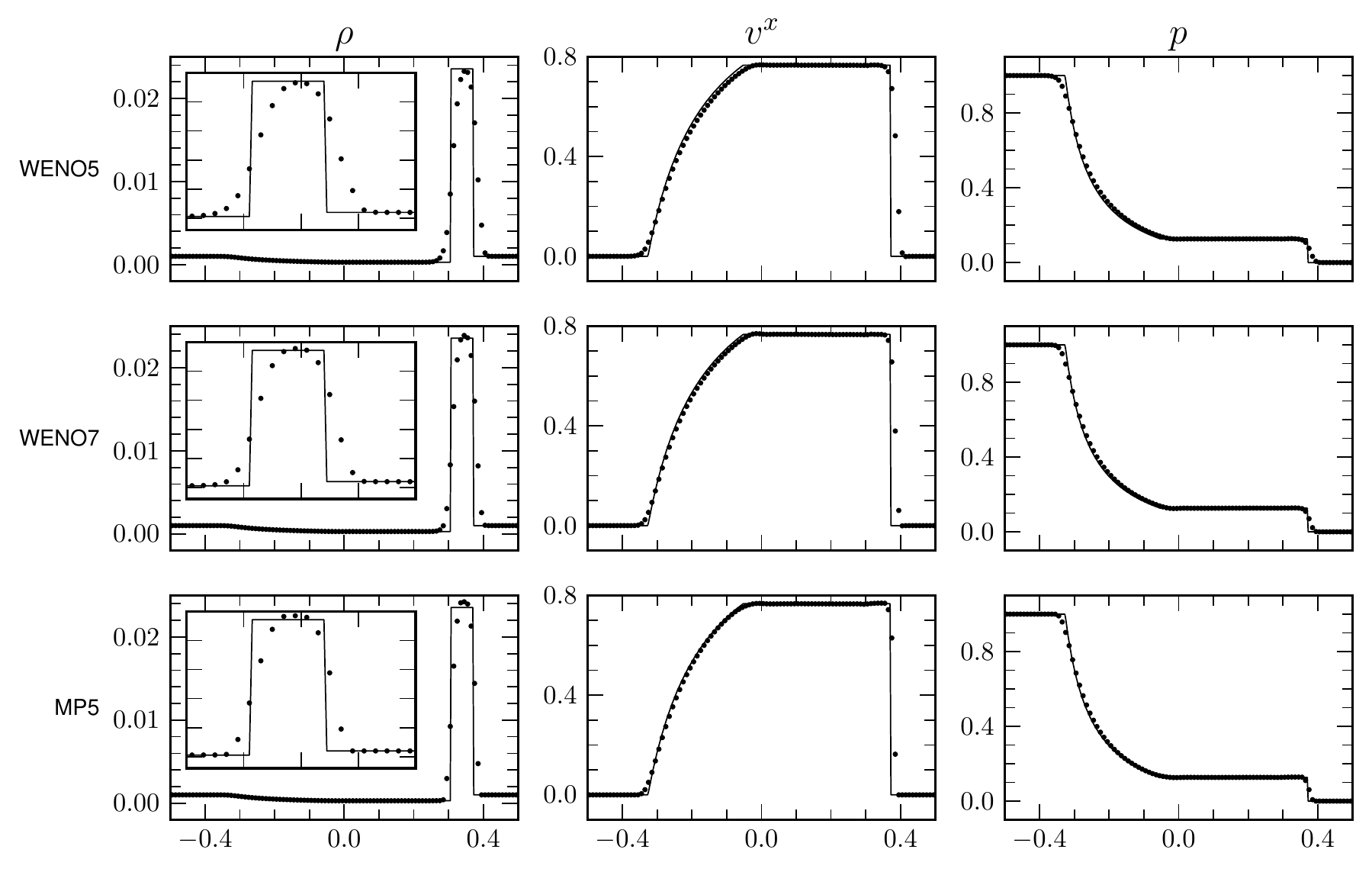}
  \end{center}
  \caption{The same as in Figure \ref{fig:newtonian.strong.shock}, but
    for the relativistic blast-wave test. The resolution is $\Delta^1
    = 1 / 100$ and the timestep is $\Delta^0 = 0.002$ for all the
    runs.}
  \label{fig:relativistic.transverse.shock}
\end{figure*}

\subsubsection{Transverse shock}

Another peculiar difficulty of relativistic hydrodynamics and without
a Newtonian counterpart, is that the equations for the momentum in the
different directions are coupled together by the Lorentz factor: even
in one-dimensional problems the application of a transverse velocity
can change completely the solution. This feature was first pointed out
by \cite{Pons00} and \cite{Rezzolla02}, and then used by
\cite{Rezzolla03} and \cite{Aloy:2006rd} to discover new physical
effect [see also \cite{mignone_2005_ppm, Zhang2006} for a description
  of the numerical consequences of this property].

To explore the flow dynamics in this case, we consider the same
initial data as for the blast-wave test [see Section
  \ref{sec:relativistic.blastwave}], with the only difference that we
add a transverse velocity to the initial data, \ie
\begin{align}
&  v^t_L = 0\,, 
&& v^t_R = 0.99\,.
\end{align}
As discussed in \cite{Zhang2006} this is not a very challenging test,
since the velocity is added only to the cold fluid and it does not
interact with the contact discontinuity. Nevertheless it is a good test
to evaluate the capability of the code to handle the presence of a
transverse-velocity at a moderate resolution, while more extreme
configurations require a resolution which is unreasonably high for
multi-dimensional applications to obtain decent solutions. A way around
this issue is the use of adaptive-mesh-refinement (AMR)
\citep{Zhang2006}, which however would not be useful for our use of
\texttt{THC} to study of relativistic turbulence Radice \& Rezzolla in
prep.~(it is in fact debatable whether the use of AMR is advantageous in
this case).

In Figure \ref{fig:relativistic.transverse.shock} we show the analytic
and numerical solutions for this problem at time $t=0.4$. The presence
of the transverse velocity widens the state between the shock wave and
the contact discontinuity. The density contrast is also smaller with
respect to the case where no tangential velocity is
present. \texttt{THC} is able to capture the solution even at the low
resolution, $\Delta^1 = 1/100$, shown in the Figure. The MP5 scheme
overestimates slightly the density contrast, but all of the algorithms
are able to capture the correct location of the shock wave.

\begin{figure*}
  \begin{center}
    \includegraphics[width=17cm, height=10cm]{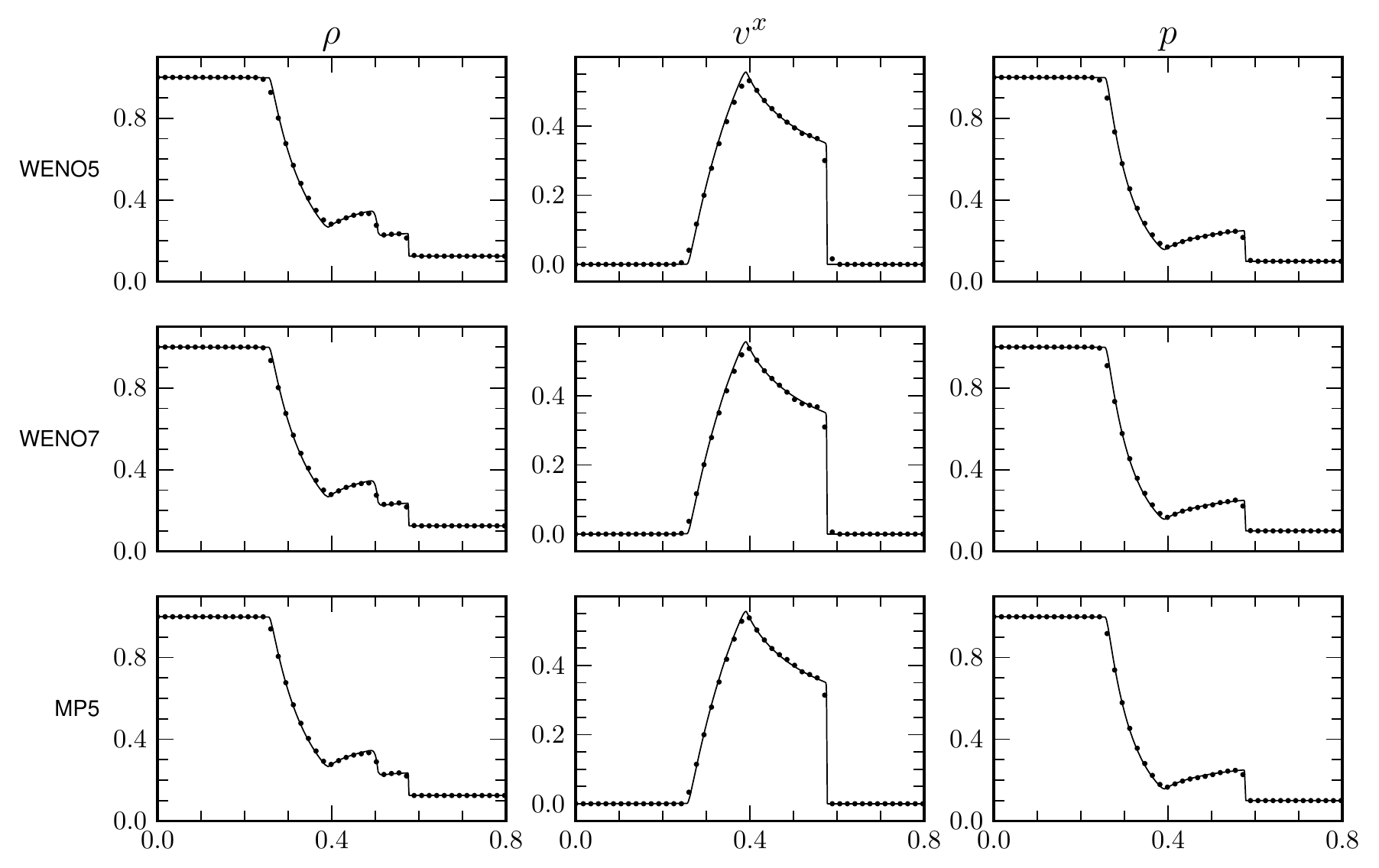}
  \end{center}
  \caption{Rest-mass density (left panels) velocity (middle panels)
    and pressure (right panels) for the relativistic spherical
    explosion test. We show the results obtained with different
    schemes (dotted) as well as a reference solution (solid line)
    obtained with the 1D spherically symmetric code \texttt{EDGES},
    \cite{Radice2011}, using 2000 elements of degree 3. The resolution
    is $\Delta^i = 1/100$ and the timestep is $\Delta^0 = 0.00125$
    for all the runs.}
  \label{fig:relativistic.explosion}
\end{figure*}

\subsubsection{Spherical explosion}

As an example of a test involving non-grid-aligned shocks we consider
the classical test of the spherical explosion in relativistic
hydrodynamics. The initial data in this case is given by
\begin{equation}
  (\rho, v, p) = \begin{cases}
    (1, 0, 1)\,, & \textrm{if } r < 0.4\,; \\
    (0.125, 0, 0.1)\,, & \textrm{elsewhere}\,.
  \end{cases}
\end{equation}
Since no analytic solution is known in this case, we use as reference
solution the one computed using the one-dimensional
spherically-symmetric discontinuous-Galerkin code \texttt{EDGES}
\citep{Radice2011} using 2000 elements of degree 3, solid lines, and
compare it with the numerical solutions obtained in three-dimensions
with \texttt{THC} in the diagonal direction. Both solutions at time $t
= 0.25$ are shown in Figure \ref{fig:relativistic.explosion}, when
using a resolution $\Delta^i = 1/100$ and a timestep $\Delta^0 =
0.00125$ for all the numerical schemes. As in the one-dimensional
case, a small timestep is necessary in order to avoid numerical
oscillations with the MP5 algorithm, while the other schemes appear to
be stable even with a timestep which is twice as large.

Overall, all the schemes are able to properly capture the reference
solution even at this very low resolution: the contact discontinuity
is captured within two grid points and no oscillations are found in
any of the physical quantities.

\begin{figure}
  \resizebox{\hsize}{!}{%
\includegraphics{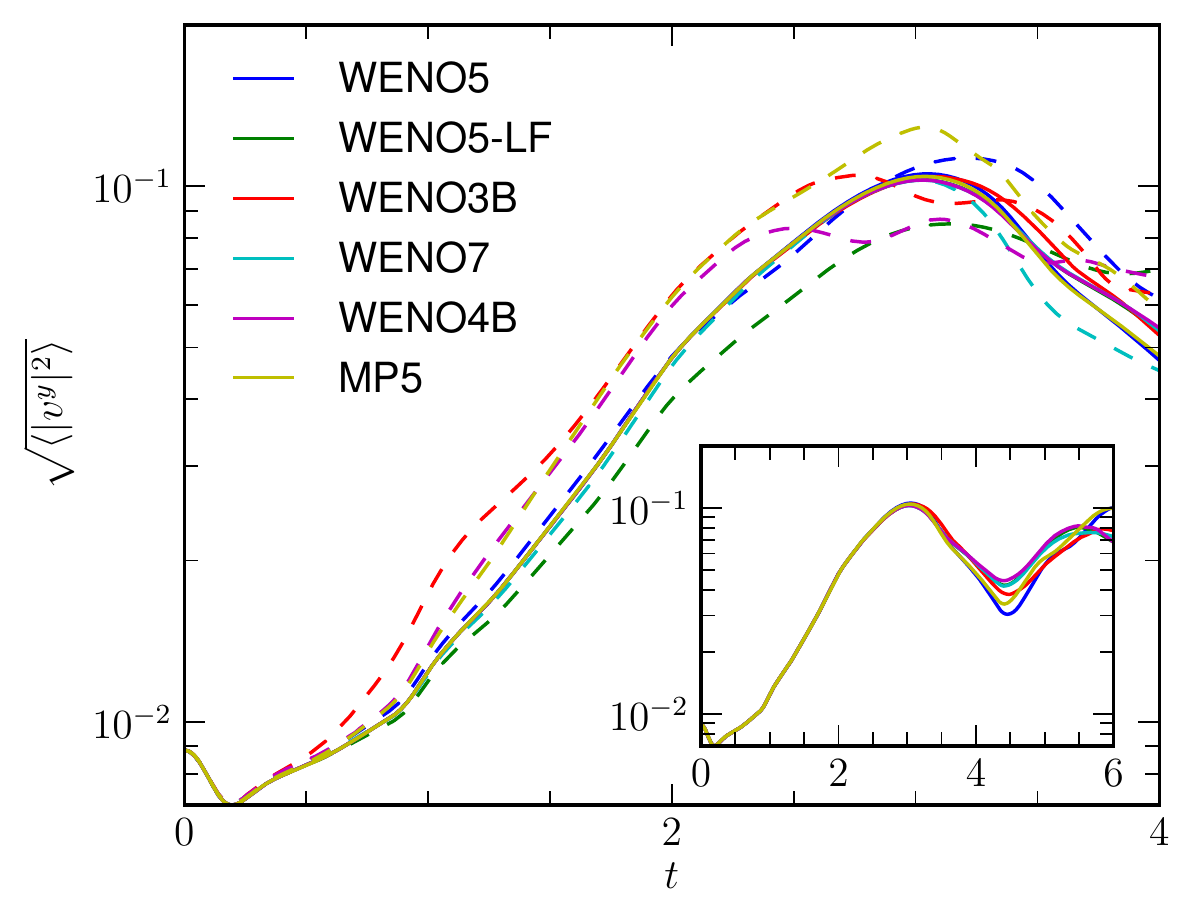}}
  \caption{RMS value of the $y$-component of the velocity during the
  linear-growth phase of the 2D relativistic Kelvin-Helmholtz
  instability, for different numerical schemes. The solid lines represent
  the results obtained at the highest resolution, $1024\times 2048$,
  while the dashed lines represent the results obtained at the lowest
  resolution, $128 \times 256$. The timestep is chosen so that the CFL
  factor is $C \approx 0.25$ for all of the runs.}
  \label{fig:relativistic.kh.vely}
\end{figure}

\begin{figure*}
  \begin{center}
    \includegraphics[width=17cm]{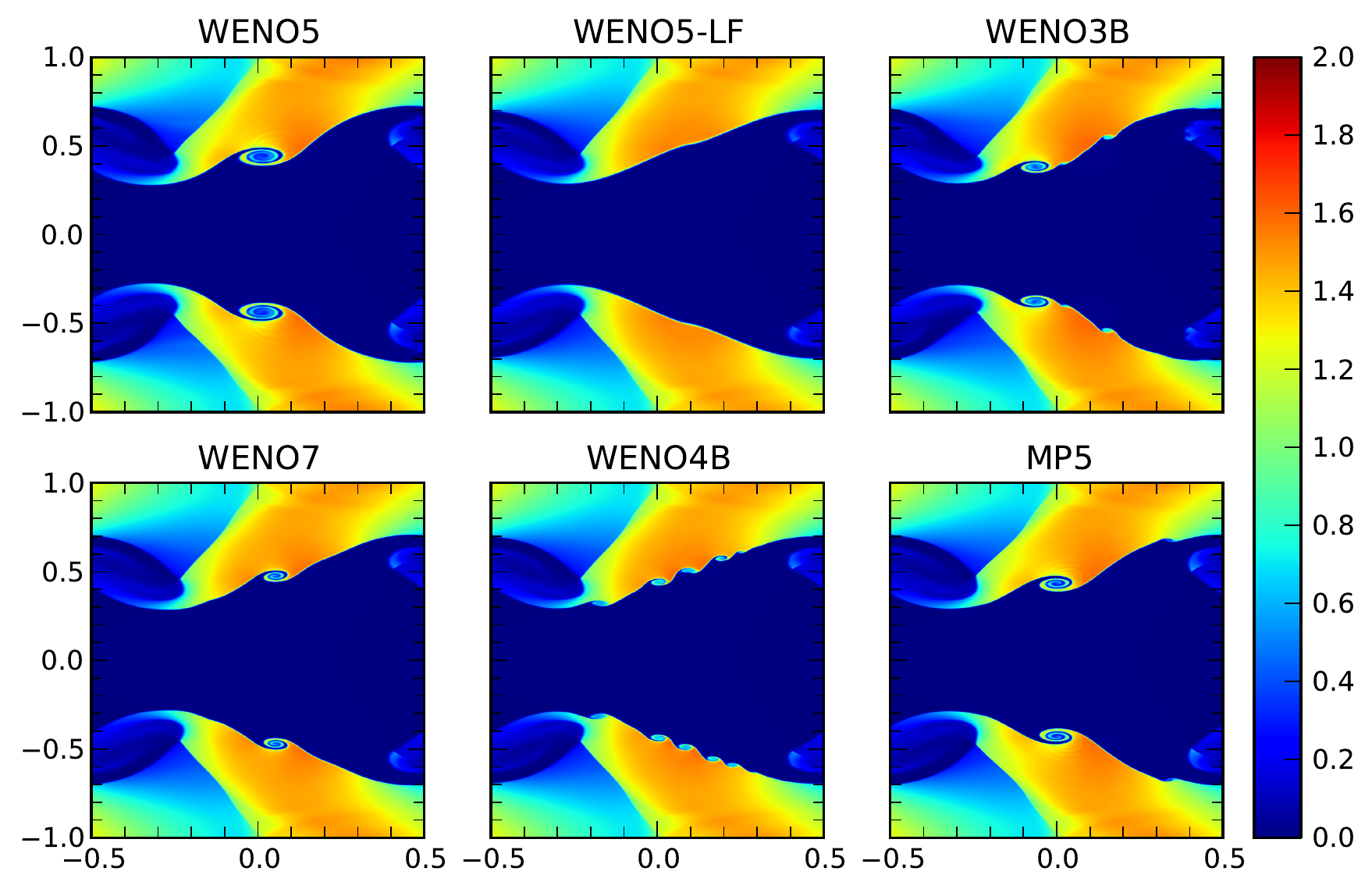}
    \caption{Rest-mass density for the 2D relativistic
      Kelvin-Helmholtz instability test at time $t=3$ and for
      different numerical schemes. The resolution is $512\times 1024$
      for all the schemes and the CFL factor is $C\approx 0.25$. This
      figure should be compared with its equivalent
      Figure~\ref{fig:relativistic.kh.rho.1024} presented in
      Appendix~\ref{sec:app_a} and at a resolution $1024\times 2048$}
    \label{fig:relativistic.kh.rho.512}
  \end{center}
\end{figure*}

\subsubsection{Kelvin-Helmholtz instability in 2D}
\label{sec:kh2d}

The last test that we present is a classical benchmark for
multidimensional hydrodynamics codes: the simulation of the KHI in
two-dimensions, $x, y$. In order to ease the comparison with the
existing literature, the initial conditions are chosen following
\cite{Beckwith2011}. The shear velocity is given by
\begin{equation}
  v^x(y) = \begin{cases}
    V_{\mathrm{shear}} \tanh\left[(y-0.5)/a\right]\,,
      & \textrm{if } y > 0\,; \\
    V_{\mathrm{shear}} \tanh\left[(y+0.5)/a\right]\,,
      & \textrm{if } y \leq 0\,;
  \end{cases}
\end{equation}
where $a = 0.01$ is the characteristic size of the shear layer and
$V_{\mathrm{shear}} = 0.5$, corresponding to a relative Lorentz
factor, \ie the Lorentz factor of a part of the fluid as seen by an
observer comoving with the other one, of $W=2.29$. The instability is
seeded by adding a small perturbation in the transverse component of
the velocity,
\begin{equation}
  v^y (x,y)= \begin{cases}
    A_0 V_{\mathrm{shear}} \sin(2\pi x) \exp\left[-\left(
      {y-0.5}\right)^2/\sigma\right]\,, & \textrm{if } y > 0\,;\\
    -A_0 V_{\mathrm{shear}} \sin(2\pi x) \exp\left[-\left(
      {y+0.5}\right)^2/\sigma\right]\,, & \textrm{if } y \leq 0\,;
  \end{cases}
\end{equation}
where $A = 0.1$ is the perturbation amplitude and $\sigma=0.1$ its
characteristic lengthscale. The adiabatic constant is $\Gamma = 4/3$ and
the initial pressure is uniform, $p = 1$. The rest-mass density
distribution, which is uniform in the $x$-direction, is instead given
by
\begin{equation}
  \rho (y) = \begin{cases}
    \rho_0 + \rho_1 \tanh\left[(y-0.5)/a\right]\,,
      & \textrm{if } y > 0\,; \\
    \rho_0 - \rho_1 \tanh\left[(y+0.5)/a\right]\,,
      & \textrm{if } y \leq 0\,;
  \end{cases}
\end{equation}
where $\rho_0 = 0.505$ and $\rho_1 = 0.495$, so that $\rho = 1$ in the
regions with $v^x = 0.5$ and $\rho = 0.1$ in the regions with $v^x =
-0.5$. Finally the computational domain is $-0.5 \leq x \leq 0.5$, $-1
\leq y \leq 1$ and we use periodic boundary conditions in all the
directions. Differently from \cite{Beckwith2011} we do not add a
random perturbation to the initial data and we do not take into
account the effects of magnetic fields. Nevertheless, our results are
in good agreement with the ones by \cite{Beckwith2011} in the linear
phase of the instability, since the magnetic field that they use is
too weak to play a dynamical role in this phase.

We performed a series of simulations using six different numerical
schemes: WENO5, WENO7, MP5, the bandwidth optimized WENO3B, WENO4B and
using the WENO5 scheme, but with the Lax-Friedrichs flux-split,
WENO5-LF. For each of these schemes, we ran with four different
resolutions: $128 \times 256$, $256 \times 512$, $512 \times 1024$ and
$1024 \times 2048$. For all the runs, the CFL factor was taken to be
$C \approx 0.25$.

The first quantity of interest to check is the growth rate of the
transverse velocity during the linear-growth phase of the KHI, as
computed with the different numerical schemes. This can be done by
comparing the evolution of the root-mean-square (RMS) value of the
transverse component of the velocity and defined as 
\begin{equation}
\langle |v^y|^2 \rangle \equiv \frac{1}{V}\int_V |v^y|^2\, dV \,,
\end{equation}
where $V$ is the volume of the computational domain. This is shown in
Figure \ref{fig:relativistic.kh.vely}, where, for each numerical scheme,
we show the results taken at the lowest resolution (dashed lines) and at
the highest one (continuous lines). First, we notice that with our setup
the linear-growth phase of the instability lasts up to until $t \simeq
3$. At that time, in fact, the transverse velocity reaches saturation and
afterwards the fluid starts to become turbulent and the velocity shows
large fluctuations.

\cite{Mignone2009} showed the importance of including the contact wave in
the approximate Riemann solver in the case of a finite-volume code. In
particular, they observed that while the growth rate of $v^y$ was already
accurate at low resolution when using their HLLD Riemann solver, in the
cases where the more dissipative HLLE Riemann solver was employed the
correct growth rate was recovered only at high resolution [similar
results were also reported by \cite{Beckwith2011}]. In analogy with what
is observed with finite-volume codes, we also remark the importance of
avoiding excessive dissipation in the contact discontinuity by comparing
the results obtained with the Lax-Friedrichs and the Roe flux-split when
using WENO5. The Lax-Friedrichs flux-split underestimates the growth rate
at low resolution and achieves a good accuracy in its measure only at
resolution $256\times 512$ (not shown in the figure). WENO5 and
WENO7 with the Roe flux split already have a growth rate which is
consistent with the highest resolution runs at the lowest resolution of
$128 \times 256$.

The behaviour of the MP5 scheme, as well as the one of the
bandwidth-optimized WENO schemes, is more surprising: all of these
schemes overestimate the growth of the RMS transverse velocity at low
resolution. This problem disappears as we increase the resolution and
in the $256\times 512$ the growth rate is already consistent with the
highest-resolution one for all the numerical schemes.

Some insight about the numerical viscosity can be gained by looking at
the topology of the flow during the linear-growth phase of the
KHI. In particular in Figure \ref{fig:relativistic.kh.rho.512} we show a
colour map of the rest-mass density obtained with the different schemes
at time $t = 3$ using the $512\times 1024$ resolution. \cite{Mignone2009}
noticed that their solution obtained with the HLLD Riemann solver was
showing a different structure, with the development of secondary
small-scale instabilities along the shear layer, and that were not
present when using the more diffusive HLLE solver. Similar differences
were also observed in other works [see, \eg \cite{Agertz2007,
Springel2010}], and the presence (or absence) of these secondary
instabilities has been often used as an indication of the numerical
viscosity of the codes. \cite{Beckwith2011} interpret these differences
as an indication that HLLE is converging to a different weak solution
from HLLD. Since both solvers are entropic (\ie with non-decreasing
entropy), this would imply the non-uniqueness of the entropic solution
for the Euler equations in this case.

In agreement with the conclusions of \cite{Beckwith2011}, we also note
that these secondary instabilities, although only numerical artifacts
(see below), appear only in schemes able to properly treat the initial
contact discontinuity. As a result, the rest-mass density in Figure
\ref{fig:relativistic.kh.rho.512} obtained with WENO5 and WENO5-LF match
very well the ones they obtained using HLLD and HLLE, respectively.
However, our results do not support their conjecture that different
schemes are converging to different weak solutions of the equations. The
reason is that these secondary instabilities appear not to be genuine
features of the solution and, rather, tend to disappear as the resolution
is increased. For instance, the number of secondary vortices seems to
change in a non-predictable fashion with the different numerical schemes
and also with the resolution. This can be seen in Figure
\ref{fig:relativistic.kh.rho.512}, which shows the great variability in
the solutions obtained with different schemes. A similar variability is
also present in results obtained at different resolutions with each
scheme, WENO5-LF being the only exception. Finally, we point out that the
size of the vortices is also shrinking as the resolution is increased.
In essence, therefore, all of these considerations lead us to the
conclusion that the secondary instabilities are triggered by the
nonlinear dissipation mechanism of the different schemes, emerge neatly
when computed with numerical schemes that treat properly the initial
contact discontinuity, but do not have a physical meaning (see also the
discussion in Appendix~\ref{sec:app_a}).

A similar interpretation is also given by \cite{McNally2011},
who go one step further and suggest to add additional viscosity to
numerical codes displaying these secondary instabilities in order to
prevent their growth. While the addition of extra dissipation is going to
smooth out small scales numerical perturbations, we argue that this issue
can only be resolved with the inclusion of physical viscosity. Artificial
viscosity would probably compete with the internal, non-linear,
dissipation mechanism of the schemes yielding results that would be even
more difficult to interpret (as the viscous scale will now depend both on
the resolution and on the artificial viscosity strength).

\begin{figure}
  \resizebox{\hsize}{!}{%
\includegraphics{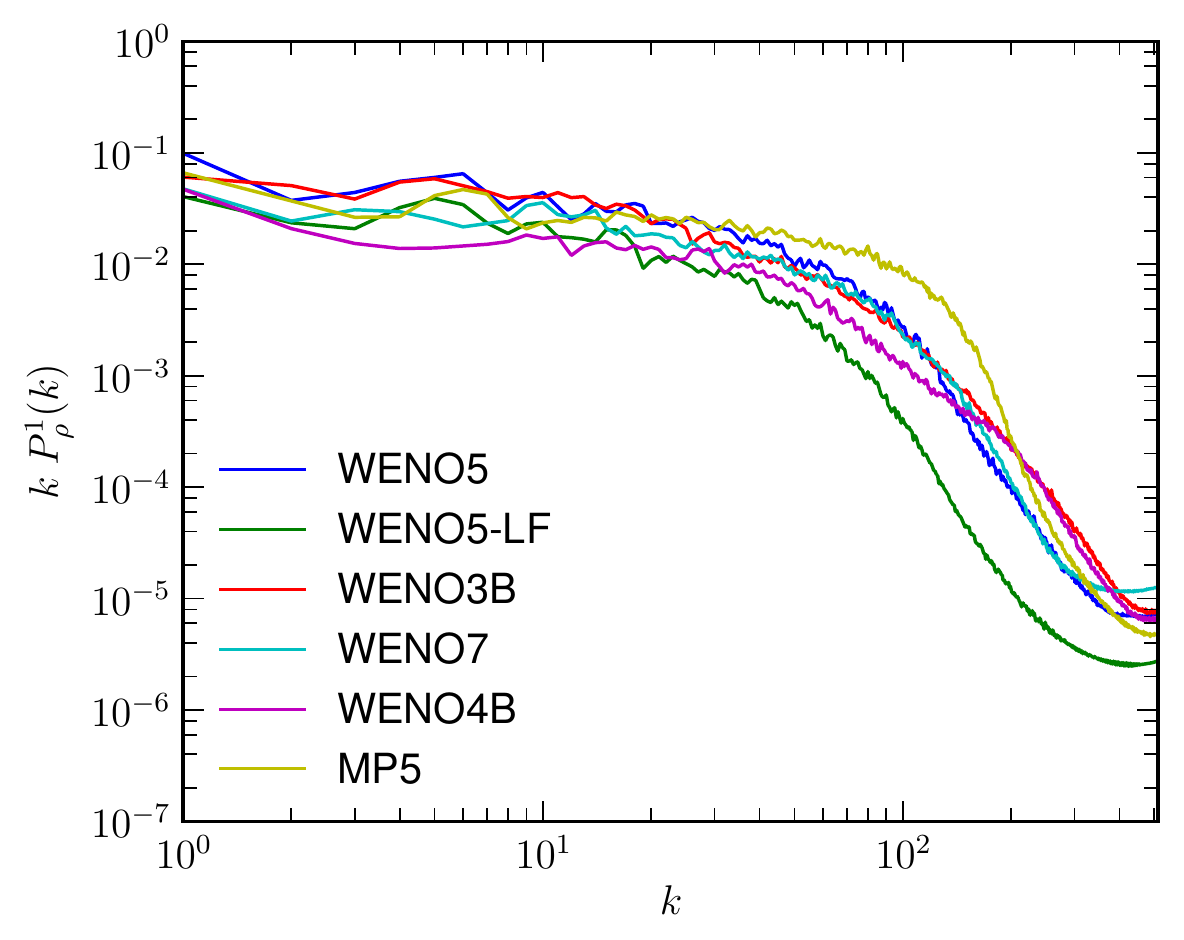}}
  \caption{One dimensional power spectrum of the rest-mass density for
    the 2D relativistic Kelvin-Helmholtz instability test at time
    $t=3$ and for different numerical schemes. The resolution is
    $1024\times 2048$ for all the schemes and the CFL factor is
    $C \approx 0.25$.}
  \label{fig:relativistic.kh.rho.spectrum}
\end{figure}
%
%
%

A more quantitative way of estimating the numerical viscosity of the
code in this test has been proposed by \cite{Beckwith2011} and is
based on the study of one-dimensional integrated power-spectra. Given a
quantity $u(x,y)$ we define its integrated power-spectrum
\begin{equation}
  P^1_u(k) = \int_{-1}^1 |\hat{u}(k,y)|^2 d y\,,
\end{equation}
where $k$ is the wavenumber and
\begin{equation}
  \hat{u}(k, y) = \int_{-1/2}^{1/2} u(x,y) e^{-2\pi i k x} d x\,,
\end{equation}
is the one-dimensional Fourier transform of $u$. To ease the
comparison with the spectra reported by \cite{Beckwith2011}, the power
spectrum is then normalized so that
\begin{equation}
  \sum_{k=1}^{N/2} P^1_u(k) = 1\,,
\end{equation}
where $N$ is the number of gridpoints. The one-dimensional power
spectrum can be used to quantify the typical scale of structures, such
as the secondary vortices discussed above, stretched in the direction
of the bulk shear flow.

In Figure \ref{fig:relativistic.kh.rho.spectrum} we show the spectra
of the rest-mass density for the different schemes at time $t = 3$ and
for the highest resolution. As expected the WENO5-LF scheme stands out
as the scheme having the largest dissipation. More surprising is the
behaviour of the bandwidth optimized schemes: they appear to be
improving over their classical counterparts only at high wavenumbers,
that is, at scales that are dominated by the (non-physical)
dissipation of the scheme. Even more unexpected is the ability of the
MP5 scheme to resolve small scales structures and that, on the basis
of the argument about the development of the secondary instabilities,
should be more dissipative than WENO4B, but which instead appears to
yield more small-scale structures in the rest-mass density.

A similar conclusion can be obtained by studying the spectrum of the
kinetic energy
\begin{equation}
  E_k = \rho W ( W - 1 )\,,
\end{equation}
but which we do not report since its behaviour is very similar to the
one shown for the rest-mass density.

\section{The relativistic Kelvin-Helmholtz instability in 3D}
\label{sec:kh3d}

As an example of a non-trivial application of \texttt{THC} and a
perfect playground to evaluate the performances of the different
numerical schemes for the simulation of turbulence reported
in Radice \& Rezzolla in prep., we present here a study of the relativistic
turbulence generated by the KHI in three-dimensions. Our analysis is
not meant to be a systematic assessment of the accuracy of these
methods for direct numerical simulations of compressible relativistic
turbulence, as done, for instance by \cite{Kitsionas2009}, or
\cite{johnsen_2010_ahr} in the case of classical turbulence. Rather,
our analysis is meant to assess how the different methods reproduce
the same turbulent initial-value problem and to provide some insight
on the spectral properties of the different schemes.

The relativistic KHI [see, \eg \cite{Bodo2004}] is of particular
interest because of its relevance for the stability of relativistic
jets [see, \eg \cite{Perucho2007, Perucho2010}], and because of its
potential role in the amplification of magnetic fields in gamma-ray
bursts [see, \eg \cite{Zhang09}], and binary neutron-star mergers
\citep{Baiotti08, Giacomazzo:2009mp, Obergaulinger10, Rezzolla:2011}.

We consider a triply-periodic box, $-0.5 \leq x \leq 0.5$, $-1 \leq y
\leq 1$, $-0.5 \leq z \leq 0.5$. The initial conditions are the same
ones employed in Section \ref{sec:kh2d}, and the symmetry in the
$z$-direction is broken with the application of random perturbations,
uniformly distributed in the range $[0, 0.01]$, applied to the
velocity in the $z$-direction $v^z$. We simulated this system up to
the time $t = 30$ on a grid of $256 \times 512 \times 256$ cells using
the WENO5, WENO7, MP5, WENO3B and WENO4B reconstructions. In addition,
we have also performed one run at the same resolution using the
second-order MINMOD reconstruction. Furthermore, because of the high
computational costs involved to check the convergence of the code, we
have managed to run only one model at twice the reference resolution,
\ie $512 \times 1024 \times 512$, using the WENO5 reconstruction
scheme.

As discussed in the previous Section, the MP5 scheme requires a
timestep which is half that of the other schemes. On the other hand,
there is no need to run the WENO schemes with such a small timestep
and in any ``real'' application we will run the WENO scheme with the
maximum possible timestep. For this reason, and since we are
interested in the performance of the different schemes in their
``real-world'' configuration,, we have not used the same CFL factor
for all of them, with a considerable saving in computational costs. In
particular, for all the runs we have used a CFL factor $C \approx
0.25$, with the exception of the one with the MP5 scheme, where we
have used $C \approx 0.125$.

\begin{figure}
  \resizebox{\hsize}{!}{%
\includegraphics{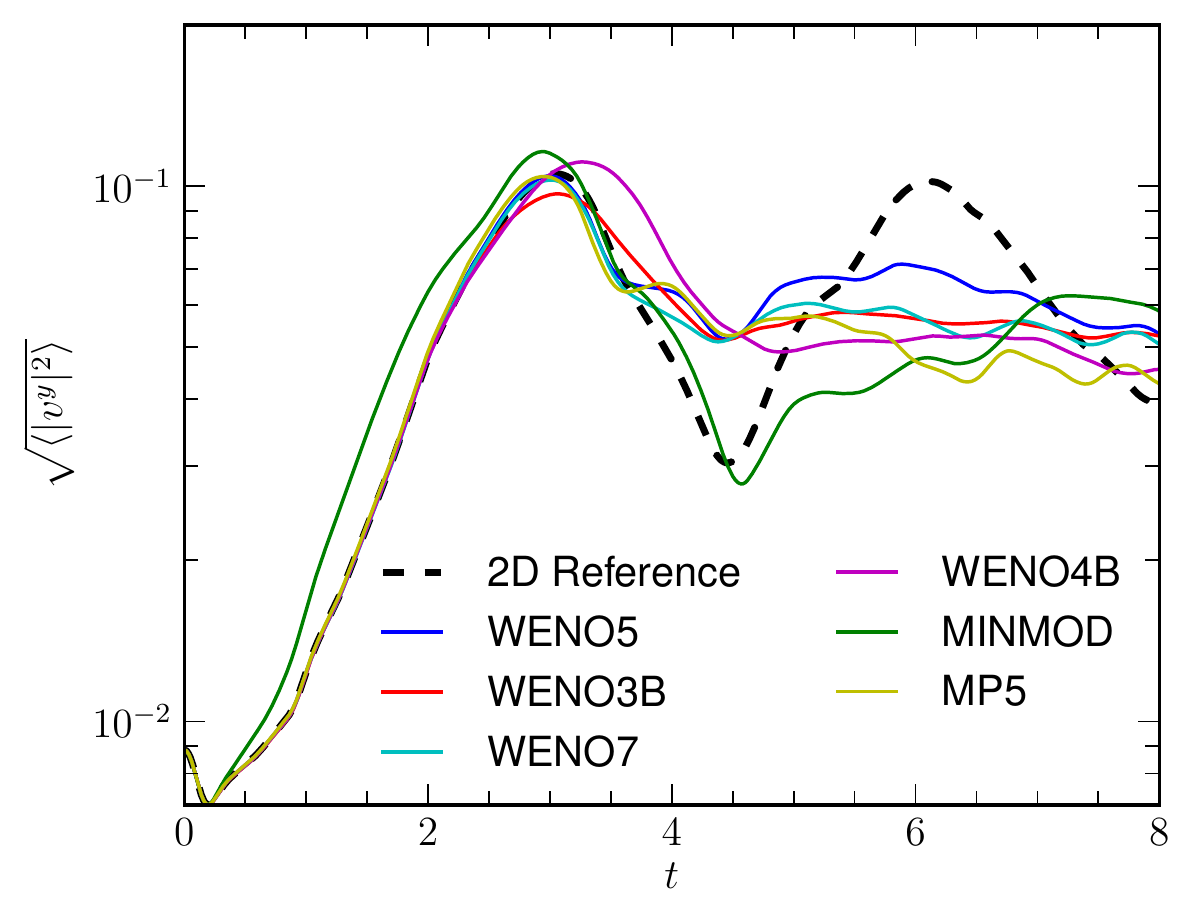}}
  \caption{RMS value of the $y$-component of the velocity during the
  linear-growth phase of the 3D relativistic Kelvin-Helmholtz
  instability, for different numerical schemes. The thick-dashed line
  represent the results obtained at the highest resolution, $1024\times
  2048$, in 2D with WENO5. The resolution is $256\times 512 \times 256$
  for all the schemes.}
  \label{fig:relativistic.kh.3d.vely}
\end{figure}

\subsection{The linear evolution of the instability}

First of all, we consider the evolution of the instability during its
linear-growth phase. At this stage, the velocity perturbations
in the direction perpendicular to the shearing one is growing
exponentially and three-dimensional effects are still negligible.

The evolution of the RMS value of the $y$-component of the velocity is
reported in Figure \ref{fig:relativistic.kh.3d.vely}, where we show the
results obtained with the different schemes, as well as a reference
solution computed in two dimensions (2D) using WENO5 on $1024\times 2048$
grid points. As expected, all the numerical schemes, with the exception
of MINMOD, are in very good agreement with the 2D solution up to the end
of the linear-growth phase at time $t = 3$, when 3D effects
become important and turbulence starts to play an important role in the
dynamics. It is interesting to note that MINMOD, which is the most
dissipative of the schemes we are using, is actually overestimating the
growth of the KHI. This suggests that some care should be taken when
interpreting the results from under-resolved simulations, since it is not
always safe to assume that the high numerical viscosity of the
low-resolution runs tends to suppress the instability, yielding a
lower-bound on its growth.

\begin{figure*}
  \begin{center}
    \includegraphics[width=17cm]{%
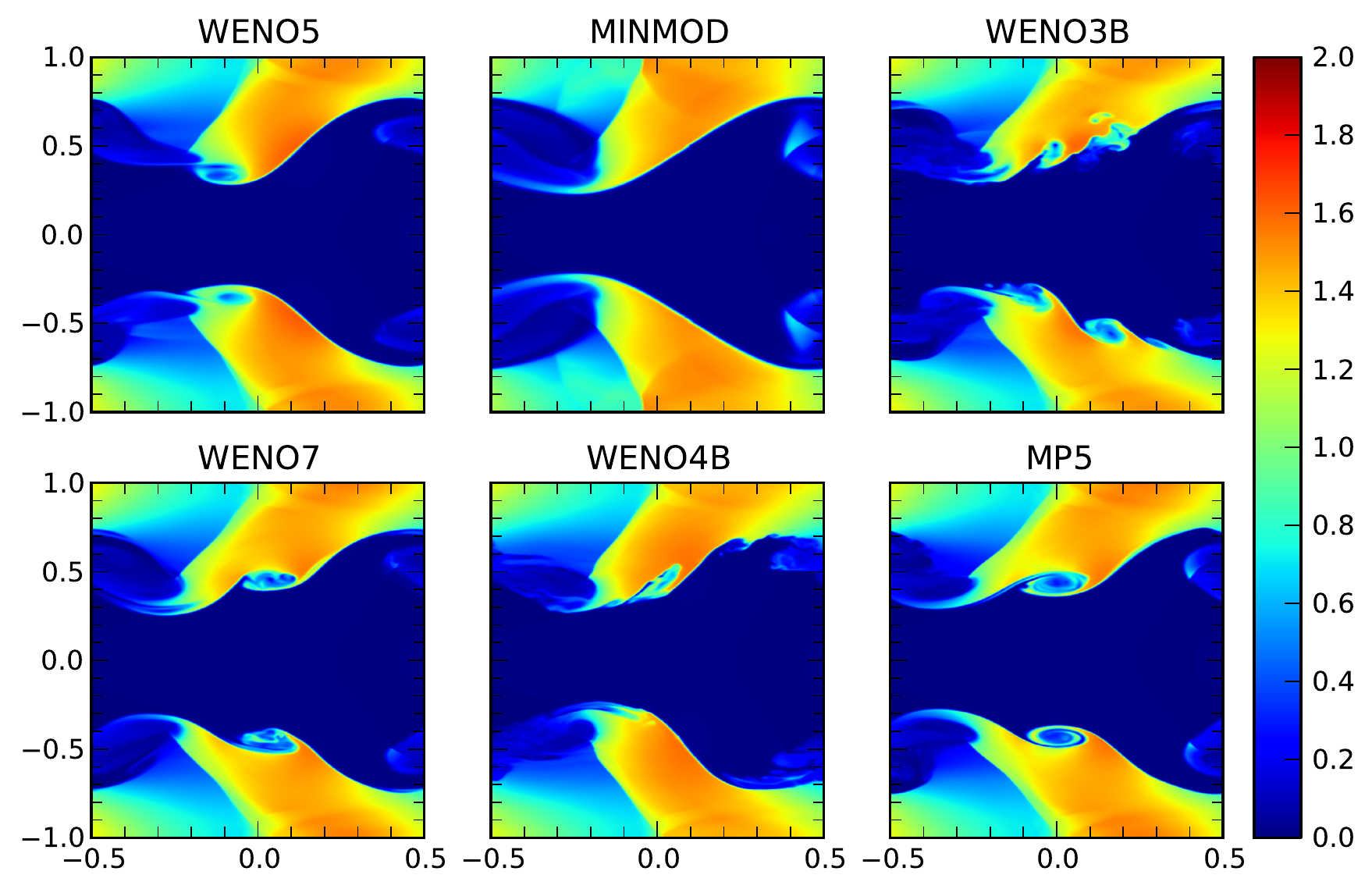}
    \caption{Rest-mass density in the $z=0$ plane for the 3D
      relativistic Kelvin-Helmholtz instability study at time $t=3$
      and for different numerical schemes. The resolution is
      $256\times 512 \times 256$ for all the schemes.}
    \label{fig:relativistic.kh.3d.rho.256}
  \end{center}
\end{figure*}

The rest-mass density at time $t = 3$ is shown in Figure
\ref{fig:relativistic.kh.3d.rho.256}, where, in analogy with the 2D
case, we find that the more dissipative schemes (\ie MINMOD), do not
show any sign of secondary instabilities apart from the seeded ones,
while the least dissipative ones (\ie WENO3B, WENO4B) show the
emergence of secondary vortices.

At this point in time, the flow is still mostly two-dimensional, but
it is interesting to notice the effects resulting from the small
perturbations in the vertical velocity $v^z$. The perturbations have
the same statistical properties in all the different models, but their
effects are appreciably different for the different schemes, as can be
seen from Figure \ref{fig:relativistic.kh.3d.rho.256}. Although the
perturbation in $v^z$ is random and it does not preserve any of the
symmetries of the problem, it is still symmetric, on average, with
respect to the $y$-axis. For this reason, one does not expect to find
a large symmetry violation until the time when these perturbations
have had enough time to grow and the dynamics has started to become
dominated by three-dimensional effects. Yet, the optimized schemes
WENO3B, WENO4B appear to be much more sensitive to the symmetry
breaking than the others. The reason for this is probably that the
bandwidth optimized-algorithms appear to trigger smaller-scale
secondary instabilities, which, in turn, are more easily affected by
the perturbations in the vertical velocity, since the perturbation
does not average out at small scales.

\begin{figure*}
  \begin{center}
    \includegraphics[width=17cm]{%
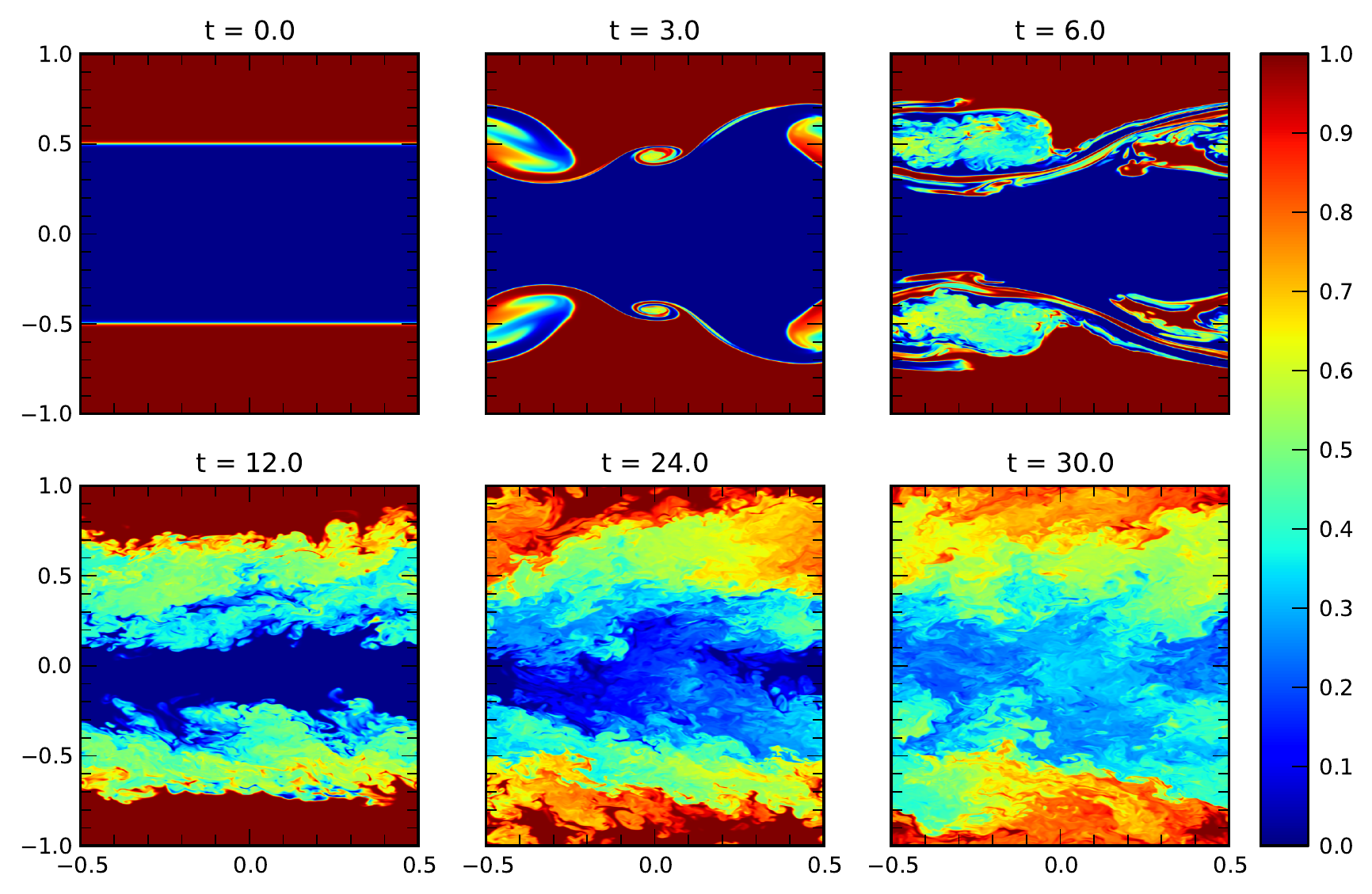}
    \caption{Evolution of the concentration of the passive tracer in
      the $z=0$ plane for the 3D Kelvin-Helmholtz instability obtained
      using WENO5 on $512 \times 1024 \times 512$ gridpoints.}
    \label{fig:relativistic.kh.3d.tracer.512}
  \end{center}
\end{figure*}

\subsection{The nonlinear evolution of the instability}

The linear-growth phase of the KHI instability ends when the
primary vortices become unstable to secondary instabilities and the flow
starts the transition to turbulence. At this point, three-dimensional
effects start to become dominant and the dynamics is qualitatively
different from the one observed in the 2D case.

In order to better track the evolution of the fluid in this phase, we
monitor the concentration of a passive scalar field, $\phi$,
transported with the fluid via the advection equation
\begin{equation}
\label{eq:tracer}
  \frac{\partial \phi}{\partial t} +
  \sum_{i=1}^3 v^i \frac{\partial \phi}{\partial x^i} = 0 \,.
\end{equation}
Equation~\eqref{eq:tracer} is not a conservation law, so that in can
not be written directly\footnote{Equation (\ref{eq:tracer}) can be
  written in conservation form at the price of introducing an
  additional, conserved variable, $\tilde{\phi} \equiv \rho W
  \phi$. In our study this complication is not necessary, as we use
  the tracer only as a diagnostic quantity. On the other hand, in
  situations where, for instance, the tracers are used to model the
  chemical composition of a fluid in a reacting flow, it may be
  important to ensure the conservation of the different species and a
  conservative approach may be preferred [see, \eg \cite{Plewa1999}
    for an example of this approach].} in the form (\ref{eq:claw}),
but it is nevertheless a hyperbolic equation of the Hamilton-Jacobi
type. For this reason, in order to solve numerically (\ref{eq:tracer})
we use a class of numerical schemes designed for this family of
equations and that can be built using the non-oscillatory
reconstruction of the derivative introduced in Section \ref{sec:code}
[see, \eg \cite{Shu2007}]. In particular, the semi-discrete form of
equation (\ref{eq:tracer}) becomes
\begin{equation}
  \begin{split}
  \frac{d \phi_{i,j,k}}{d t} =&
      v^x_{i,j,k} \frac{\phi_{i-1/2,j,k} - \phi_{i+1/2,j,k}}{\Delta^1} +
      v^y_{i,j,k} \frac{\phi_{i,j-1/2,k} - \phi_{i,j+1/2,k}}{\Delta^2} + \\
    & v^z_{i,j,k} \frac{\phi_{i,j,k-1/2} - \phi_{i,j,k+1/2}}{\Delta^3}\,,
  \end{split}
\end{equation}
where $(\phi_{i-1/2,j,k} - \phi_{i+1/2,j,k}) / \Delta^1$ is a
high-order non-oscillatory approximation of $-\partial \phi / \partial
x^1$ at $x_{i,j,k}$ obtained using an upwind-biased reconstruction,
\ie
\begin{equation}
\phi_{i-1/2,j,k} = \phi^{\pm}_{i-1/2,j,k}, \qquad \textrm{if } v^x \lessgtr 0\,.
\end{equation}
The ``fluxes'' in the other directions are obviously computed in an
analogous way.

The time evolution of the tracer, as computed with WENO5 at the
highest-resolution is shown in Figure
\ref{fig:relativistic.kh.3d.tracer.512}. At the initial time, we set
the scalar field $\phi$ to be equal to the rest-mass density, so that
the initial configuration consists in two regions with opposite
``colour'', separated by a thin transition layer. At the end of the
linear regime, \ie at time $t \simeq 3$, the tracer profile is
distorted by the presence of the primary vortices as well as of the
secondary ones, but there is only a marginal mixing of the two
``phases'' of the fluid. Note that, since we do not include any
explicit dissipation term in the advection equation (\ref{eq:tracer}),
the mixing of the tracer happens only due to the numerical
dissipation. As the vortices start to become unstable, the fluid
starts to concentrate the scalar field in thin structures and the two
regions of the fluid start to mix around the shearing region. As the
simulation proceeds, the width of the region effected by the mixing
grows, till at time $t \simeq 24$, when there are only small patches
of the fluid that still have a uniform tracer (colour). At the final
time, $t = 30$, there are no regions in the flow that are unaffected
by the mixing and the initial structure is mostly destroyed, even
though perfect mixing has not been achieved yet.

We track the evolution of the variance of the scalar field, which we
compute as
\begin{equation}
 \mathrm{Var}[\phi] \equiv 
\langle |\phi - \langle\phi\rangle|^2\rangle\,,
\end{equation}
and which, for $t \geq 5$, we find to be very-well fitted by a simple
exponential law of the type
\begin{equation}
  \mathrm{Var}[\phi] = K e^{-t/\tau}\,.
\end{equation}
The values of the fitting constant for the WENO5 scheme at the highest
resolution are $K = 0.28$ and $\tau = 14.75$. The mixing timescale,
$\tau$, exhibits only small changes between the runs at different
resolutions, with the exception of the results obtained with MP5, where
the timescale is $\tau = 17.5$. Hence, the total time of the
simulation, $t=30$, is roughly equivalent to two e-folding times for
the mixing of the passive tracer.

\subsubsection{Fully-Developed turbulence}

After the linear-growth phase, the flow quickly becomes
turbulent. By the time we stop the simulation at $t = 30$, the turbulence
is fully developed even though the flow is still not isotropic, but
preserving some memory of the initial configuration. Because of the large
computational costs involved in these tests (which, we recall, have been
performed for six different methods) we have not carried out the
simulations for longer times, assuming that the properties of the
non-perfectly isotropic turbulence at the final time is sufficiently
close to the fully isotropic turbulence. We point out that our simulation
time is more then two times longer than the one reported in
\cite{Zhang09}, where a setup similar to ours was used, but in the more
challenging regime of relativistic MHD.
%
%
\begin{figure}
  \resizebox{\hsize}{!}{\includegraphics{%
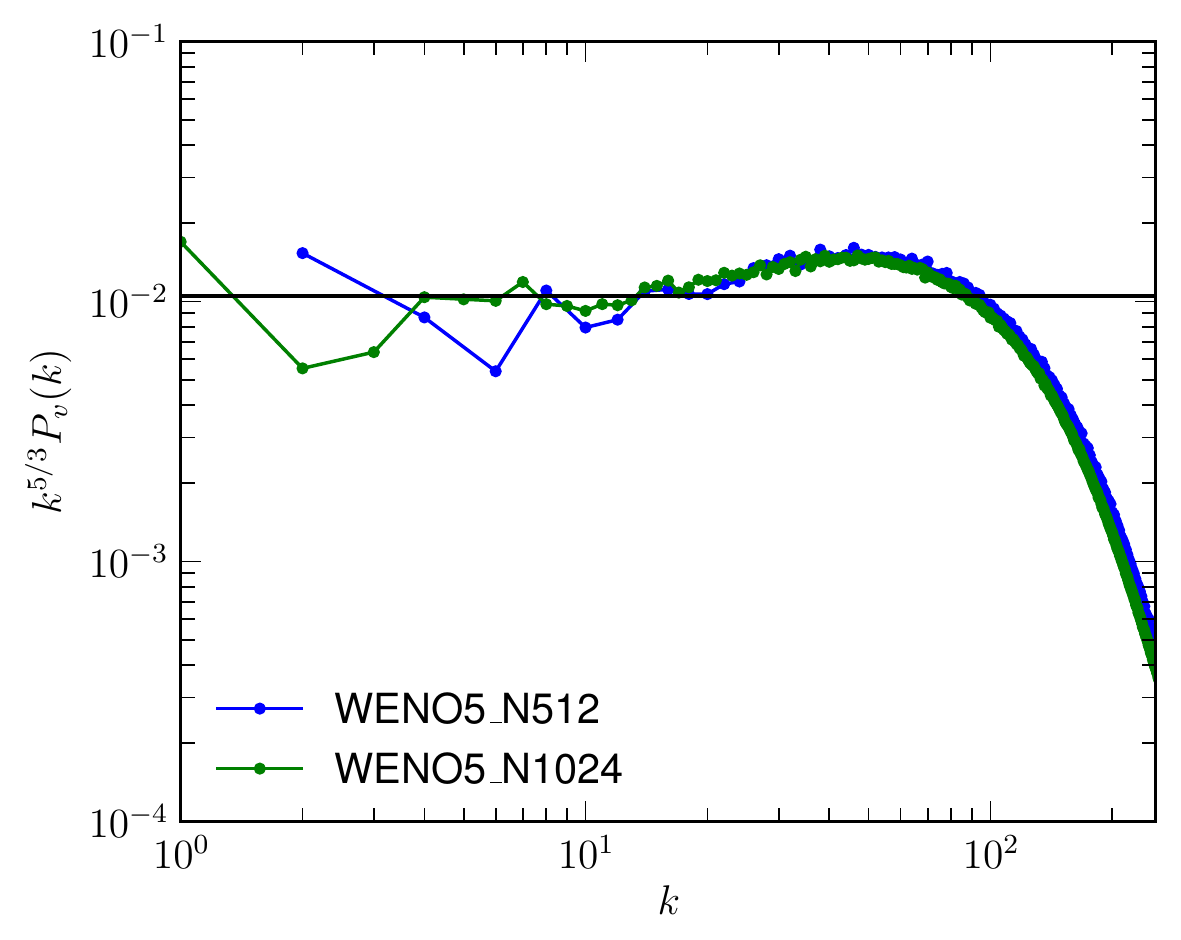}}
  \caption{Compensated power spectrum of the velocity at time $t=30$ for the 3D
  Kelvin-Helmholtz instability test, using WENO5 at two different resolutions,
  $256 \times 512 \times 256$ and $512 \times 1024 \times 512$.}
  \label{fig:relativistic.kh.3d.vel.spectrum.convergence.scaled}
\end{figure}
\begin{figure*}
  \begin{center}
    \includegraphics{%
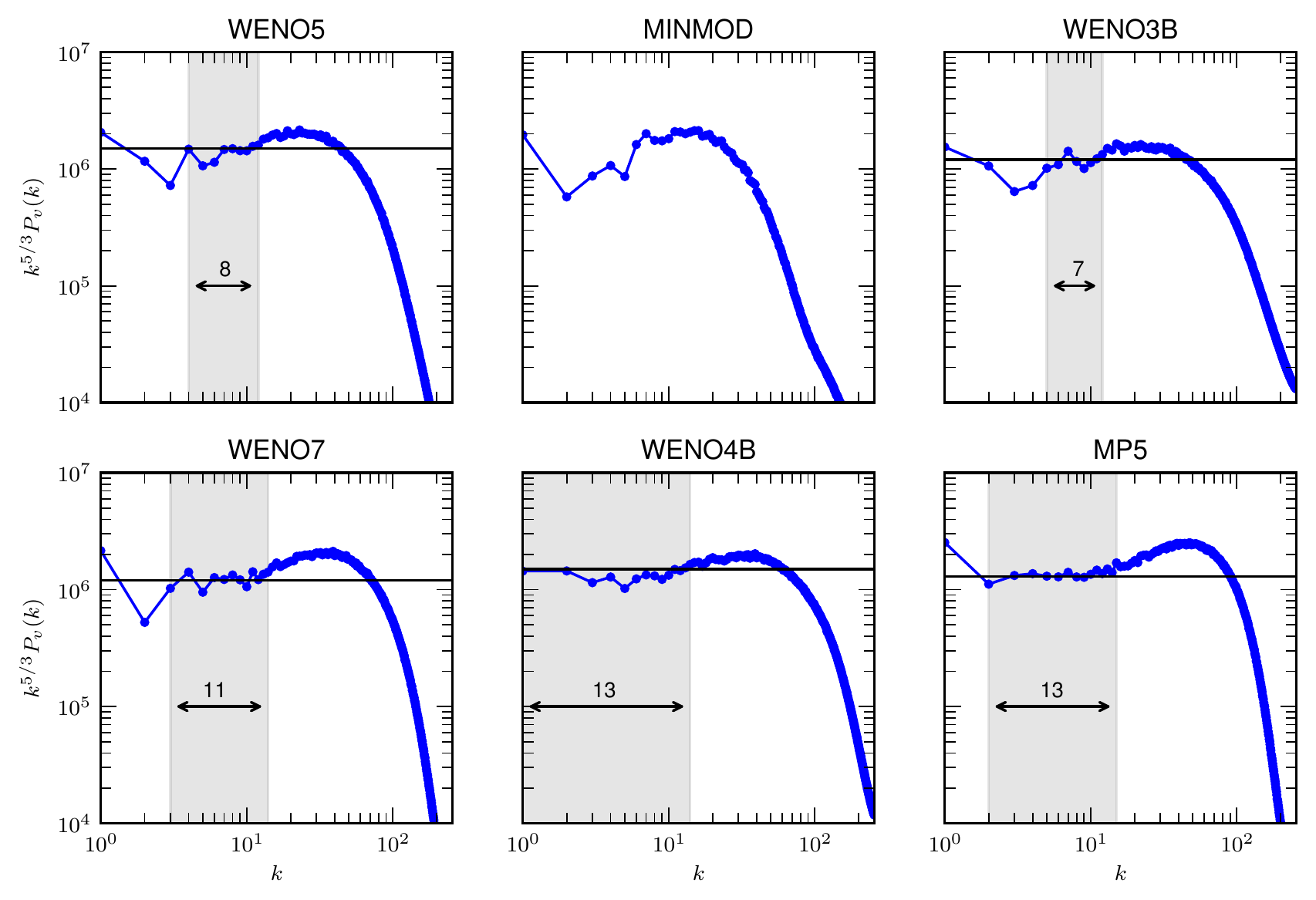}
    \caption{Compensated power spectrum of the velocity at time $t=30$ for the
    3D Kelvin-Helmholtz instability test, using different numerical schemes. In
    all the cases the resolution is $256\times 512 \times 256$.}
    \label{fig:relativistic.kh.3d.vel.spectrum.comparison}
  \end{center}
\end{figure*}

By far the most interesting quantity to study is the three-dimensional
velocity power spectrum
\begin{equation}\label{eq:spectrum}
  P_v(k) = \frac{1}{2} \int_{|\boldsymbol{k}|=k} |\hat{v}(\boldsymbol{k})|^2\,
    d \boldsymbol{k}\,,
\end{equation}
where $\hat{v}(\boldsymbol{k})$ is the three-dimensional Fourier transform of
$v(\boldsymbol{x})$,
\begin{equation}
  \hat{v}(\boldsymbol{k}) = \int_V v(\boldsymbol{x})
    e^{-2\pi i \boldsymbol{k}\cdot \boldsymbol{x}} d \boldsymbol{x}\,,
\end{equation}
and $V$ is the volume of the computational region. The integral in
(\ref{eq:spectrum}) is computed following \cite{Eswaran1988} as
\begin{equation}
  P_v(k) = \frac{1}{2}\frac{4 \pi k^2}{N_k}
    \sum_{k-1/2 < |\boldsymbol{k}| \leq k+1/2} |\hat{v}(\boldsymbol{k})|^2\,,
\end{equation}
where $N_k$ is the number of discrete wave-numbers in the shell $k-1/2
\leq |\boldsymbol{k}| < k+1/2$. For simplicity, we study the data in
the artificially-enlarged cubic domain $[-1,1]^3$ and we do that by
duplicating original data in the $x$ and $z$-directions, exploiting
the symmetry of the problem. This procedure avoids the complications
of a computational domain which does not have the same extent in all
directions and yields an ``equivalent resolution'' of $512^3$ and
$1024^3$ points for the low and high-resolution runs
respectively. Clearly the first few wave-numbers of the spectrum are
influenced by this procedure, but all the higher wave-numbers are
essentially unaffected.

At the final time, the flow is subsonic ($\mathcal{M}_{\mathrm{max}}
\lesssim 1$) and relativistically warm ($\epsilon \gtrsim 1$). Under
these conditions, studies of Newtonian \citep{Porter2002} and
relativistic \citep{Zhang09, Inoue2011, Zrake2011a}
transonic turbulence suggest that the velocity spectrum should be
consistent with the Kolmogorov phenomenology
\citep{Kolmogorov1991a}. In particular, the power spectrum should
scale as
\begin{equation}
  P_v(k) \sim k^{-5/3}\,,
\end{equation}
in the so called \emph{inertial range}, that is at those scales where
the fluid motion is sufficiently independent from the large-scale
dynamics and from the small-scale viscosity, so as to exhibit a
selfsimilar universal behaviour.

In Figure \ref{fig:relativistic.kh.3d.vel.spectrum.convergence.scaled}
we show the compensated velocity spectrum, \ie $k^{5/3} P_v(k)$, at
time $t = 30$ obtained from the data of the two WENO5 runs. More
specifically, the low-resolution spectrum is shifted to larger
wavenumbers by a factor two and scaled assuming a $k^{-5/3}$ scaling
of the spectrum. The rationale behind this procedure is that we are
interested in the (eventual) selfsimilar behaviour of the spectrum and
it is therefore useful to consider the low-resolution run as a
realisation of the same flow as the high-resolution one, but in a
smaller volume. In other words, we interpret the spectrum of the
low-resolution run as being computed from a subsample of the data of
the high-resolution one [see, \eg \cite{Bodo2011} for a more detailed
  discussion of the issue of convergence for direct simulations of
  turbulent flows].

The plot demonstrates the statistical convergence of the simulation,
with the exception of the very low-wavenumber part of the spectrum,
where the convergence is probably spoiled by the symmetry. At the same
time, the plot also shows that only the high-resolution run seems to
be able to cover a sufficiently wide part of the inertial range to
give a clear indication of the Kolmogorov $-5/3$ scaling.

In Figure \ref{fig:relativistic.kh.3d.vel.spectrum.comparison} we show
the compensated velocity spectra at time $t = 30$ obtained with the
different numerical schemes. For each scheme, we highlight with a grey
shaded area the part of the spectrum that appears to be ``compatible''
with the $-5/3$ scaling inferred from the high-resolution WENO5
run. The width of this region, expressed in terms of wavenumbers, is
also indicated on the figure. Clearly, this measure has to be taken
with a bit of caution, since there is a certain degree of
arbitrariness in the identification of the ``inertial range'';
furthermore, the judgment is also made more difficult by the fact that
all of the results are obtained at a resolution which probably is not
high-enough and that a convergence study is only available for the
WENO5 case. Notwithstanding these caveats, the difference between the
various schemes is already sufficiently clear to deduce some of their
properties despite the limited amount of data that we were able to
generate.

A first conclusion to be drawn is about the importance of the use of
high-order schemes, whch is apparent if we compare the different
spectra with the one obtained with the MINMOD scheme. This
second-order algorithm, in fact, yields a spectrum which is completely
dominated by the so-called ``bottleneck-effect'' [see, \eg
  \cite{Brandenburg2011, Schmidt2006}], \ie by a region where the
power-spectrum shows an excess due to viscous effects. No clear
indication of an inertial range appears anywhere in the spectrum, with
the only ``flat'' region being the middle of the bottleneck zone. This
could be easily mistaken as the inertial range in the absence of a
proper convergence analysis. For this reason, and as remarked many
times already, care should be taken while interpreting the results
obtained with low-order schemes.

Secondly, we observe that WENO4B has an effective resolution which is
about $50\%$ larger then the one from WENO5 and $20\%$ larger then
WENO7.  Given that saving a factor $1.5$ in resolution corresponds, in
3D to a decrease of the computation costs by a factor five\footnote{We
  note that when increasing the resolution, also the timestep is
  reduced via the CLF condition. Hence, the additional cost goes like
  the fourth power of the ratio in resolutions.}, we conclude that the
use of WENO4B over WENO5 is well justified, since WENO4B is roughly
twice as expensive as WENO5 in 3D. On the other hand, the spectral
resolution of WENO4B does not appear to be better than the one of the
MP5 scheme, which has similar computational costs (due to the stricter
CFL limitation), but which is also expected to have better parallel
scaling because of its more compact stencil. Overall, MP5 shows an
``inertial-range'' as large as WENO4B. We also note that WENO3B does
not seem to yield any improvements over WENO5.

All things considered, the main differences between the
bandwidth-optimized schemes and their traditional counterparts seem to
lay in the bottleneck region: WENO3B and WENO4B have a much less
pronounced bottleneck with respect to WENO5, WENO7 and MP5. This suggests
that these schemes should be considered especially for under-resolved
turbulent flows, where the spectrum is basically entirely dominated by
the dissipation. MP5, on the other hand, can be particularly useful for
those problems where one is interested in well-resolved quantities, such
as in direct numerical simulations of turbulence, since the scales
affected by the numerical dissipation are more easily identified by the
pronounced bottleneck. MP5 should also be the scheme of choice in
Newtonian hydrodynamics, since there its timestep constraint seems to be
less severe.

\section{Conclusions}\label{sec:conclusions}

We have presented \texttt{THC}, a new multi-dimensional,
finite-difference, high-resolution shock-capturing code for classical
and special-relativistic hydrodynamics. \texttt{THC} employs up to
seventh-order accurate reconstruction of the fluxes in local
characteristic variables and the Roe flux-vector-splitting algorithm
with a novel entropy-fix prescription. The multi-dimensional case is
treated in a dimensionally unsplit fashion and the time integration is
done with a third-order strongly-stability-preserving Runge-Kutta
scheme.

We have carried out a systematic comparison of the results obtained
with our code using five different reconstruction operators: the
classical WENO5, WENO7, MP5, as well as two specialised
bandwidth-optimized WENO schemes: WENO3B and WENO4B. For all schemes,
we have checked their ability to sharply capture grid-aligned,
diagonal or spherical shock waves, and have carried out a rigorous
assessment of their accuracy in the case of smooth solutions. Finally,
we have contrasted the performance of the different methods in the
resolution of small scale structures in turbulent flows. To the best
of our knowledge, this is the first time that such a comparison has
been carried over in the special-relativistic case.

Among the different tests studied, some are highly nontrivial, such those
involving the linear and the nonlinear phases of the development of the
Kelvin-Helmholtz instability for a relativistic fluid in two and three
dimensions. In particular, we have shown the importance of using schemes
that are able to properly capture the initial contact discontinuity in
order to obtain the correct growth rate of the RMS transverse velocity in
the linear-growth phase of the instability at low resolution,
confirming the findings by \cite{Mignone2009} and \cite{Beckwith2011}.

When studying Kelvin-Helmholtz instability in two dimensions, we have
investigated the nature of the secondary vortices that appear during
the initial stages of the instability when using some of the numerical
schemes considered. We have then clarified that these vortices are not
genuine features of the solution of the equations, but rather
numerical artefacts that converge away with resolution. When studying
Kelvin-Helmholtz instability in three dimensions, we have instead
investigated the ``mixing timescale'' of the instability and the
subsequent turbulent flow, showing that we are able to obtain a
converged measure of the velocity power spectrum, using the WENO5
scheme. Our data offers a clear indication that the Kolmogorov
phenomenology \citep{Kolmogorov1991a} holds also for the turbulence in
a subsonic relativistically-warm fluid. Using the Kolmogorov $-5/3$
scaling as a reference, we have estimated the effective inertial range
of the different schemes, highlighting the importance of using
high-order schemes to study turbulent flows. The code has also been
used in a systematic investigation of direct numerical simulations of
driven turbulence in an ultrarelativistic hot plasma, whose results
will be presented in a companion paper (Radice \& Rezzolla in prep.).

Finally, \texttt{THC} represents the first step towards the
implementation of new and high-order methods for the accurate study of
general-relativistic problems, such as the inspiral and merger of
binary neutron stars~\citep{Baiotti2011} and their relation with
gamma-ray bursts~\citep{Rezzolla:2011}. We are in fact convinced that
the transition towards higher-order methods is now a necessary and an
inevitable step for a more realistic description of the complex
phenomenology associated with these relativistic-astrophysics
processes.

\begin{acknowledgements}
We thank Petros Tzeferacos for the help in the coding of the MP5
reconstruction and Wolfgang Kastaun for the support in the implementation
of the primitive recovery routine used in \texttt{THC} and for useful
comments on the draft of this work. We are also grateful to Aaryn Tonita,
Filippo Galeazzi, Ian Hawke, Colin P.~McNally, Andrew MacFadyen
and Andrea Mignone for discussions.
\end{acknowledgements}

\appendix
\section{Secondary vortices in the KHI}
\label{sec:app_a}

\begin{figure*}
  \begin{center}
    \includegraphics[width=17cm]{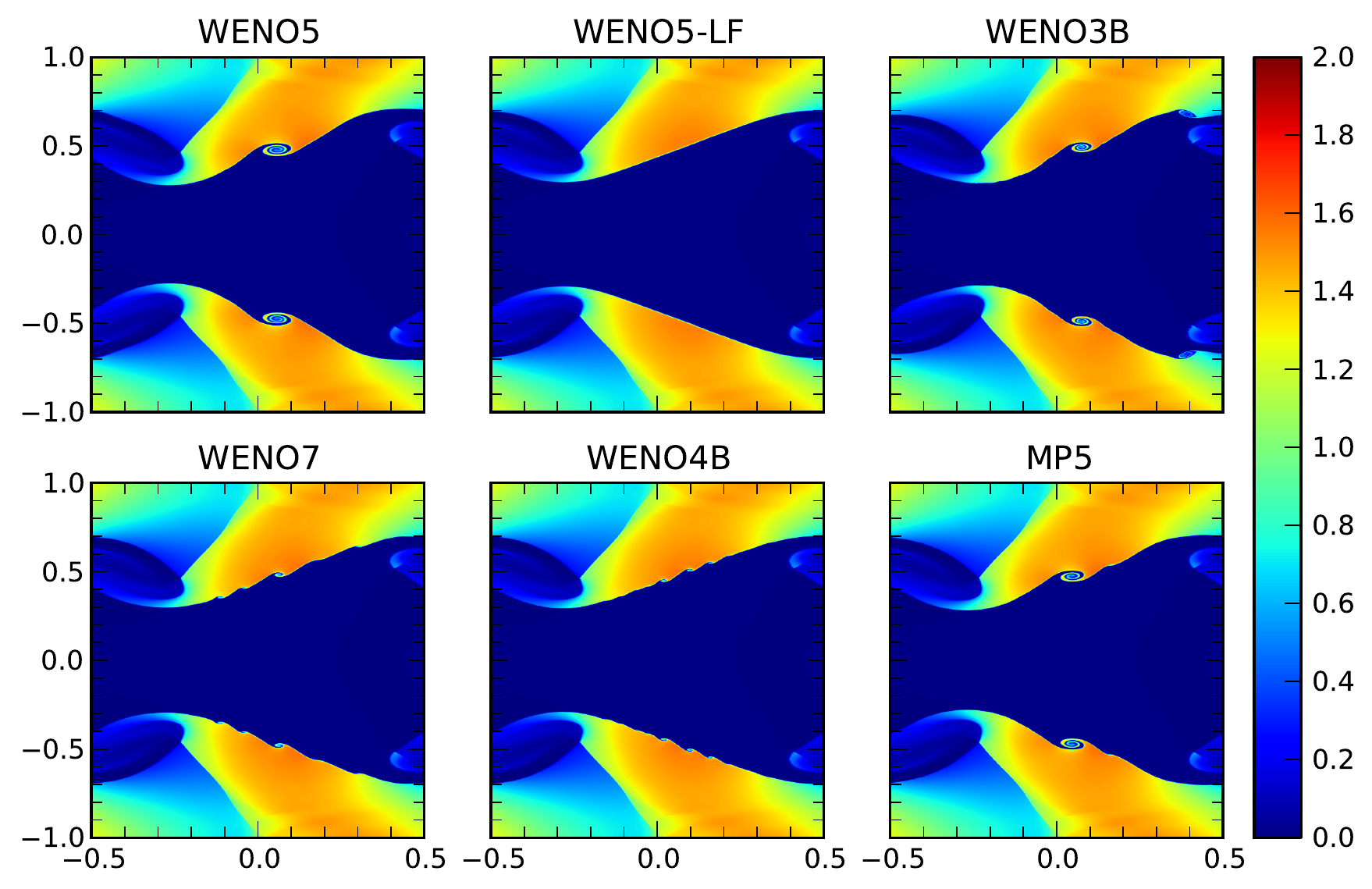}
    \caption{Rest-mass density for the 2D relativistic
      Kelvin-Helmholtz instability test at time $t=3$ and for
      different numerical schemes. The resolution is $1024\times 2048$
      for all the schemes and the CFL factor is $C\approx 0.25$. This
      figure should be compared with its equivalent
      Figure~\ref{fig:relativistic.kh.rho.512} at a resolution
      $512\times 1024$}
    \label{fig:relativistic.kh.rho.1024}
  \end{center}
\end{figure*}

In Sect.~\ref{sec:kh2d} we have commented that we believe the secondary
instabilities appearing the in initial stages of the KHI are not genuine
features of the solution, but just artefacts of the (low) resolution. The
latter acts differently on different schemes, leading to a
non-predictable change in the number and shape of these secondary
vortices. Convincing evidence that these are indeed artifacts is shown in
Figure~\ref{fig:relativistic.kh.rho.1024}, which is the same as Figure
\ref{fig:relativistic.kh.rho.512}, but at the higher resolution of
$1024\times 2048$. When comparing the two figures it appears clear that
these secondary instabilities are much smaller.  Interestingly,
therefore, the more dissipative scheme WENO5 with the Lax-Friedrichs
flux-split, WENO5-LF, seems to converge more rapidly to the correct
solution (at least in these initial stages) than its less diffusive
counterparts.

\bibliographystyle{aa.bst}
\bibliography{aeireferences.bib,local.bib}

\begin{thebibliography}{97}
\expandafter\ifx\csname natexlab\endcsname\relax\def\natexlab#1{#1}\fi

\bibitem[{Agertz {et~al.}(2007)Agertz, Moore, Stadel, Potter, Miniati, Read,
  Mayer, Gawryszczak, Kravtsov, Nordlund, Pearce, Quilis, Rudd, Springel,
  Stone, Tasker, Teyssier, Wadsley, \& Walder}]{Agertz2007}
Agertz, O., Moore, B., Stadel, J., {et~al.} 2007, Monthly Notices of the Royal
  Astronomical Society, 380, 963

\bibitem[{Aloy {et~al.}(1999)Aloy, Pons, \& Ib{\'a}{\~n}ez}]{Aloy99a}
Aloy, M.~A., Pons, J.~A., \& Ib{\'a}{\~n}ez, J.~M. 1999, Comput. Phys. Commun.,
  120, 115

\bibitem[{Aloy \& Rezzolla(2006)}]{Aloy:2006rd}
Aloy, M.~A. \& Rezzolla, L. 2006, Astrophys. J., 640, L115

\bibitem[{{Anile}(1990)}]{Anile_book}
{Anile}, A.~M. 1990, {Relativistic Fluids and Magneto-fluids} (Cambridge
  University Press)

\bibitem[{Ant{\'o}n {et~al.}(2006)Ant{\'o}n, Zanotti, Miralles, Mart{\'\i},
  Ib{\'a}{\~n}ez, Font, \& Pons}]{Anton06}
Ant{\'o}n, L., Zanotti, O., Miralles, J.~A., {et~al.} 2006, Astrophys. J., 637,
  296

\bibitem[{Baiotti {et~al.}(2010)Baiotti, Damour, Giacomazzo, Nagar, \&
  Rezzolla}]{Baiotti:2010}
Baiotti, L., Damour, T., Giacomazzo, B., Nagar, A., \& Rezzolla, L. 2010, Phys.
  Rev. Lett., 105, 261101

\bibitem[{{Baiotti} {et~al.}(2011){Baiotti}, {Damour}, {Giacomazzo}, {Nagar},
  \& {Rezzolla}}]{Baiotti2011}
{Baiotti}, L., {Damour}, T., {Giacomazzo}, B., {Nagar}, A., \& {Rezzolla}, L.
  2011, Phys. Rev. D, 84, 024017

\bibitem[{{Baiotti} {et~al.}(2008){Baiotti}, {Giacomazzo}, \&
  {Rezzolla}}]{Baiotti08}
{Baiotti}, L., {Giacomazzo}, B., \& {Rezzolla}, L. 2008, Phys. Rev. D, 78,
  084033

\bibitem[{Baiotti {et~al.}(2009)Baiotti, Giacomazzo, \&
  Rezzolla}]{Baiotti:2009gk}
Baiotti, L., Giacomazzo, B., \& Rezzolla, L. 2009, Classical Quantum Gravity,
  26, 114005

\bibitem[{Baiotti {et~al.}(2003)Baiotti, Hawke, Montero, \&
  Rezzolla}]{Baiotti03a}
Baiotti, L., Hawke, I., Montero, P., \& Rezzolla, L. 2003, in Computational
  Astrophysics in Italy: Methods and Tools, ed. R.~Capuzzo-Dolcetta, Vol.~1
  (Trieste: MSAIt), 210

\bibitem[{Baiotti {et~al.}(2005)Baiotti, Hawke, Montero, L{\"o}ffler, Rezzolla,
  Stergioulas, Font, \& Seidel}]{Baiotti04}
Baiotti, L., Hawke, I., Montero, P.~J., {et~al.} 2005, Phys. Rev. D, 71, 024035

\bibitem[{Banyuls {et~al.}(1997)Banyuls, Font, Ib{\'a}{\~n}ez, Mart{\'\i}, \&
  Miralles}]{Banyuls97}
Banyuls, F., Font, J.~A., Ib{\'a}{\~n}ez, J.~M., Mart{\'\i}, J.~M., \&
  Miralles, J.~A. 1997, Astrophys. J., 476, 221

\bibitem[{Beckwith \& Stone(2011)}]{Beckwith2011}
Beckwith, K. \& Stone, J.~M. 2011, The Astrophysical Journal Supplement Series,
  193, 6

\bibitem[{Bernuzzi {et~al.}(2012{\natexlab{a}})Bernuzzi, Nagar, Thierfelder, \&
  Bruegmann}]{Bernuzzi2012}
Bernuzzi, S., Nagar, A., Thierfelder, M., \& Bruegmann, B. 2012{\natexlab{a}},
  arXiv:1205.3403, 12

\bibitem[{Bernuzzi {et~al.}(2012{\natexlab{b}})Bernuzzi, Thierfelder, \&
  Br\"{u}gmann}]{Bernuzzi2011}
Bernuzzi, S., Thierfelder, M., \& Br\"{u}gmann, B. 2012{\natexlab{b}}, Physical
  Review D, 85

\bibitem[{Bodo {et~al.}(2011)Bodo, Cattaneo, Ferrari, Mignone, \&
  Rossi}]{Bodo2011}
Bodo, G., Cattaneo, F., Ferrari, A., Mignone, A., \& Rossi, P. 2011, The
  Astrophysical Journal, 739, 82

\bibitem[{Bodo {et~al.}(2004)Bodo, Mignone, \& Rosner}]{Bodo2004}
Bodo, G., Mignone, a., \& Rosner, R. 2004, Phys. Rev. E, 70, 1

\bibitem[{Brandenburg \& Nordlund(2011)}]{Brandenburg2011}
Brandenburg, A. \& Nordlund, A.~k. 2011, Reports on Progress in Physics, 74,
  046901

\bibitem[{Chiu(1973)}]{Chiu1973}
Chiu, H.~H. 1973, Physics of Fluids, 16, 825

\bibitem[{Colella \& Woodward(1984)}]{colella_1984_ppm}
Colella, P. \& Woodward, P.~R. 1984, Journal of Computational Physics, 54, 174

\bibitem[{{Del Zanna} {et~al.}(2003){Del Zanna}, {Bucciantini}, \&
  {Londrillo}}]{DelZanna2003}
{Del Zanna}, L., {Bucciantini}, N., \& {Londrillo}, P. 2003, Astron.
  Astrophys., 400, 397

\bibitem[{{Del Zanna} {et~al.}(2007){Del Zanna}, {Zanotti}, {Bucciantini}, \&
  {Londrillo}}]{DelZanna2007}
{Del Zanna}, L., {Zanotti}, O., {Bucciantini}, N., \& {Londrillo}, P. 2007,
  Astron. Astrophys., 473, 11

\bibitem[{Donat {et~al.}(1998)Donat, Font, Ib{\'a}{\~n}ez, \&
  Marquina}]{Donat98}
Donat, R., Font, J.~A., Ib{\'a}{\~n}ez, J.~M., \& Marquina, A. 1998, J. Comput.
  Phys., 146, 58

\bibitem[{Duez {et~al.}(2005)Duez, Liu, Shapiro, \& Stephens}]{Duez05MHD0}
Duez, M.~D., Liu, Y.~T., Shapiro, S.~L., \& Stephens, B.~C. 2005, Phys. Rev. D,
  72, 024028, astro-ph/0503420

\bibitem[{Duffell \& MacFadyen(2011)}]{Duffell2011}
Duffell, P.~C. \& MacFadyen, A.~I. 2011, The Astrophysical Journal Supplement
  Series, 197, 15

\bibitem[{{Dumbser} \& {Zanotti}(2009)}]{Dumbser2009}
{Dumbser}, M. \& {Zanotti}, O. 2009, Journal of Computational Physics, 228,
  6991

\bibitem[{Eswaran \& Pope(1988)}]{Eswaran1988}
Eswaran, V. \& Pope, S. 1988, Computers \& Fluids, 16, 257

\bibitem[{{Farris} {et~al.}(2008){Farris}, {Li}, {Liu}, \&
  {Shapiro}}]{Farris08}
{Farris}, B.~D., {Li}, T.~K., {Liu}, Y.~T., \& {Shapiro}, S.~L. 2008, Phys.
  Rev. D, 78, 024023

\bibitem[{Font(2008)}]{Font08}
Font, J.~A. 2008, Living Rev. Relativ., 6, 4

\bibitem[{Gammie {et~al.}(2003)Gammie, McKinney, \& T{\'o}th}]{Gammie03}
Gammie, C.~F., McKinney, J.~C., \& T{\'o}th, G. 2003, Astrophys. J., 589, 458

\bibitem[{Giacomazzo \& Rezzolla(2007)}]{Giacomazzo:2007ti}
Giacomazzo, B. \& Rezzolla, L. 2007, Classical Quantum Gravity, 24, S235

\bibitem[{{Giacomazzo} {et~al.}(2009){Giacomazzo}, {Rezzolla}, \&
  {Baiotti}}]{Giacomazzo:2009mp}
{Giacomazzo}, B., {Rezzolla}, L., \& {Baiotti}, L. 2009, Mon. Not. R. Astron.
  Soc., 399, L164

\bibitem[{Goodale {et~al.}(2003)Goodale, Allen, Lanfermann, Mass{\'o}, Radke,
  Seidel, \& Shalf}]{Goodale02a}
Goodale, T., Allen, G., Lanfermann, G., {et~al.} 2003, in Vector and Parallel
  Processing -- VECPAR'2002, 5th International Conference, Lecture Notes in
  Computer Science (Berlin: Springer)

\bibitem[{Gottlieb {et~al.}(2009)Gottlieb, Ketcheson, \& Shu}]{gottlieb2009}
Gottlieb, S., Ketcheson, D., \& Shu, C.-W. 2009, Journal of Scientific
  Computing, 38, 251, 10.1007/s10915-008-9239-z

\bibitem[{Gourgoulhon(1991)}]{gourgoulhon_1991_seg}
Gourgoulhon, E. 1991, Astronomy and Astrophysics, 252, 651

\bibitem[{Inoue {et~al.}(2011)Inoue, Asano, \& Ioka}]{Inoue2011}
Inoue, T., Asano, K., \& Ioka, K. 2011, Astrophys. J., 734, 77

\bibitem[{Jiang \& Shu(1996)}]{Jiang1996}
Jiang, G.-S. \& Shu, C.-W. 1996, J. Comput. Phys, 126, 202

\bibitem[{Johnsen {et~al.}(2010)Johnsen, Larsson, Bhagatwala, Cabot, Moin,
  Olson, Rawat, Shankar, Sj{\"{o}}green, \& Yee}]{johnsen_2010_ahr}
Johnsen, E., Larsson, J., Bhagatwala, A.~V., {et~al.} 2010, Journal of
  Computational Physics, 229, 1213

\bibitem[{Kastaun(2006)}]{kastaun_2006_hrs}
Kastaun, W. 2006, Phys. Rev. D, 74, 124024

\bibitem[{Kastaun(2007)}]{Kastaun2007}
Kastaun, W. 2007, PhD thesis, University of T{\"u}bingen

\bibitem[{Kitsionas {et~al.}(2009)Kitsionas, Federrath, Klessen, Schmidt,
  Price, Dursi, Gritschneder, Walch, Piontek, Kim, a.~K.~Jappsen, Ciecielag, \&
  {Mac Low}}]{Kitsionas2009}
Kitsionas, S., Federrath, C., Klessen, R.~S., {et~al.} 2009, Astronomy and
  Astrophysics, 508, 541

\bibitem[{{Koide} {et~al.}(1999){Koide}, {Shibata}, \& {Kudoh}}]{Koide99}
{Koide}, S., {Shibata}, K., \& {Kudoh}, T. 1999, Astrophys. J., 522, 727

\bibitem[{Kolmogorov(1991)}]{Kolmogorov1991a}
Kolmogorov, A.~N. 1991, Proceedings of the Royal Society A: Mathematical,
  Physical and Engineering Sciences, 434, 9

\bibitem[{Komissarov(2004)}]{Komissarov04}
Komissarov, S.~S. 2004, Mon. Not. R. Astron. Soc., 350, 1431

\bibitem[{{Konigl}(1980)}]{Konigl1980}
{Konigl}, A. 1980, Physics of Fluids, 23, 1083

\bibitem[{Kulikovskii {et~al.}(2001)Kulikovskii, Pogorelov, \&
  Semenov}]{Kulikovskii2001}
Kulikovskii, A., Pogorelov, N., \& Semenov, A.~Y. 2001, Mathematical aspects of
  numerical solution of hyperbolic systems (Chapman \& Hall/CRC)

\bibitem[{Leveque(1992)}]{LeVeque92}
Leveque, R.~J. 1992, Numerical Methods for Conservation Laws (Basel: Birkhauser
  Verlag)

\bibitem[{{Liu} {et~al.}(1994){Liu}, {Osher}, \& {Chan}}]{Liu1994}
{Liu}, X.-D., {Osher}, S., \& {Chan}, T. 1994, Journal of Computational
  Physics, 115, 200

\bibitem[{Mann(1985)}]{Mann1985}
Mann, P. 1985, Journal of Computational Physics, 58, 377

\bibitem[{Mart{\'\i} {et~al.}(1991)Mart{\'\i}, Ib{\'a}{\~n}ez, \&
  Miralles}]{Marti91}
Mart{\'\i}, J.~M., Ib{\'a}{\~n}ez, J.~M., \& Miralles, J.~A. 1991, Phys. Rev.
  D, 43, 3794

\bibitem[{Mart{\'\i} \& M{\"u}ller(2003)}]{Marti03}
Mart{\'\i}, J.~M. \& M{\"u}ller, E. 2003, Living Rev. Relativ., 6, 7

\bibitem[{Mart{\'{i}}n {et~al.}(2006)Mart{\'{i}}n, Taylor, Wu, \&
  Weirs}]{martin_2006_bow}
Mart{\'{i}}n, M.~P., Taylor, E.~M., Wu, M., \& Weirs, V.~G. 2006, Journal of
  Computational Physics, 220, 270

\bibitem[{May \& White(1966)}]{May66}
May, M.~M. \& White, R.~H. 1966, Phys. Rev., 141, 1232

\bibitem[{McNally {et~al.}(2011)McNally, Lyra, \& Passy}]{McNally2011}
McNally, C.~P., Lyra, W., \& Passy, J.-c. 2011, arXiv:1111.1764, 24

\bibitem[{Meier(1999)}]{meier_1999_mas}
Meier, D.~L. 1999, The Astrophysical Journal, 518, 788

\bibitem[{Mignone {et~al.}(2005)Mignone, Plewa, \& Bodo}]{mignone_2005_ppm}
Mignone, A., Plewa, T., \& Bodo, G. 2005, The Astrophysical Journal Supplement
  Series, 160, 199

\bibitem[{Mignone {et~al.}(2010)Mignone, Tzeferacos, \&
  Bodo}]{mignone_2010_hoc}
Mignone, A., Tzeferacos, P., \& Bodo, G. 2010, Journal of Computational
  Physics, 229, 5896

\bibitem[{Mignone {et~al.}(2009)Mignone, Ugliano, \& Bodo}]{Mignone2009}
Mignone, a., Ugliano, M., \& Bodo, G. 2009, Monthly Notices of the Royal
  Astronomical Society, 393, 1141

\bibitem[{Neilsen {et~al.}(2006)Neilsen, Hirschmann, \& Millward}]{Neilsen2005}
Neilsen, D., Hirschmann, E.~W., \& Millward, R.~S. 2006, Classical Quantum
  Gravity, 23, S505

\bibitem[{Noh(1987)}]{noh_1987_ecs}
Noh, W. 1987, Journal of Computational Physics, 72, 78

\bibitem[{{Obergaulinger} {et~al.}(2010){Obergaulinger}, {Aloy}, \&
  {M{\"u}ller}}]{Obergaulinger10}
{Obergaulinger}, M., {Aloy}, M.~A., \& {M{\"u}ller}, E. 2010, Astronomy and
  Astrophysics, 515, A30

\bibitem[{Perucho {et~al.}(2007)Perucho, Hanasz, Mart\'{\i}, \&
  Miralles}]{Perucho2007}
Perucho, M., Hanasz, M., Mart\'{\i}, J.-M., \& Miralles, J.-A. 2007, Phys. Rev.
  E, 75, 1

\bibitem[{{Perucho} {et~al.}(2010){Perucho}, {Mart{\'{\i}}}, {Cela}, {Hanasz},
  {de La Cruz}, \& {Rubio}}]{Perucho2010}
{Perucho}, M., {Mart{\'{\i}}}, J.~M., {Cela}, J.~M., {et~al.} 2010, Astronomy
  and Astrophysics, 519, A41

\bibitem[{Plewa \& Mueller(1999)}]{Plewa1999}
Plewa, T. \& Mueller, E. 1999, Astronomy and Astrophysics Astrophysics, 342,
  179

\bibitem[{Pons {et~al.}(2000)Pons, Mart{\'\i}, \& M{\"u}ller}]{Pons00}
Pons, J.~A., Mart{\'\i}, J.~M., \& M{\"u}ller, E. 2000, J. Fluid Mech., 422,
  125

\bibitem[{Porter {et~al.}(2002)Porter, Pouquet, \& Woodward}]{Porter2002}
Porter, D., Pouquet, A., \& Woodward, P. 2002, Phys. Rev. E, 66, 1

\bibitem[{Quirk(1994)}]{Quirk1994}
Quirk, J. 1994, International Journal for Numerical Methods in Fluids, 18, 555

\bibitem[{{Radice} \& {Rezzolla}(2011)}]{Radice2011}
{Radice}, D. \& {Rezzolla}, L. 2011, Phys. Rev. D, 84, 024010

\bibitem[{{Rezzolla} {et~al.}(2011){Rezzolla}, {Giacomazzo}, {Baiotti},
  {Granot}, {Kouveliotou}, \& {Aloy}}]{Rezzolla:2011}
{Rezzolla}, L., {Giacomazzo}, B., {Baiotti}, L., {et~al.} 2011, Astrophys. J.,
  732, L6

\bibitem[{{Rezzolla} \& {Zanotti}(2002)}]{Rezzolla02}
{Rezzolla}, L. \& {Zanotti}, O. 2002, Phys. Rev. Lett., 89, 114501

\bibitem[{Rezzolla {et~al.}(2003)Rezzolla, Zanotti, \& Pons}]{Rezzolla03}
Rezzolla, L., Zanotti, O., \& Pons, J.~A. 2003, Journ. of Fluid Mech., 479, 199

\bibitem[{Rider(2000)}]{rider_2000_rwh}
Rider, W. 2000, Journal of Computational Physics, 162, 395

\bibitem[{Roe(1981)}]{Roe81}
Roe, P.~L. 1981, J. Comput. Phys., 43, 357

\bibitem[{Rosswog(2010)}]{rosswog_2010_csr}
Rosswog, S. 2010, Journal of Computational Physics, 229, 8591

\bibitem[{Schmidt {et~al.}(2006)Schmidt, Hillebrandt, \&
  Niemeyer}]{Schmidt2006}
Schmidt, W., Hillebrandt, W., \& Niemeyer, J. 2006, Computers \& Fluids, 35,
  353

\bibitem[{Schneider {et~al.}(1993)Schneider, Katscher, Rischke, Waldhauser,
  Maruhn, \& Munz}]{Schneider90}
Schneider, V., Katscher, U., Rischke, D.~H., {et~al.} 1993, J. Comput. Phys.,
  105, 92

\bibitem[{{Sekiguchi}(2010)}]{Sekiguchi2010}
{Sekiguchi}, Y. 2010, Progress of Theoretical Physics, 124, 331

\bibitem[{{Sekiguchi} {et~al.}(2011){Sekiguchi}, {Kiuchi}, {Kyutoku}, \&
  {Shibata}}]{Sekiguchi2011}
{Sekiguchi}, Y., {Kiuchi}, K., {Kyutoku}, K., \& {Shibata}, M. 2011, Phys. Rev.
  Lett., 107, 051102

\bibitem[{Shu(2007)}]{Shu2007}
Shu, C. 2007, Lecture Notes Series, Institute for Mathematical Sciences,
  National University of Singapore, 11, 47

\bibitem[{Shu(1997)}]{Shu97}
Shu, C.~W. 1997, {E}ssentially non-oscillatory and weighted essentially
  non-oscillatory schemes for hyperbolic conservation laws, Lecture notes ICASE
  Report 97-65; NASA CR-97-206253, NASA Langley Research Center, {\tt
  http://techreports.larc.nasa.gov/icase/1997/icase-1997-65.pdf}

\bibitem[{Shu(1999)}]{Shu99}
Shu, C.~W. 1999, in High-{O}rder {M}ethods for {C}omputational {P}hysics, ed.
  T.~J. Barth \& H.~Deconinck (Springer), 439--582

\bibitem[{Shu(2001)}]{Shu01}
Shu, C.~W. 2001, {H}igh order finite difference and finite volume WENO schemes
  and discontinuous Galerkin methods for CFD, Tech. Rep. ICASE Report 2001-11;
  NASA CR-2001-210865, NASA Langley Research Center

\bibitem[{Shu \& Osher(1988)}]{Shu88}
Shu, C.~W. \& Osher, S.~J. 1988, J. Comput. Phys., 77, 439

\bibitem[{Siegler \& Riffert(2000)}]{Siegler00}
Siegler, S. \& Riffert, H. 2000, The Astrophysical Journal, 531, 1053

\bibitem[{{Sod}(1978)}]{Sod1978}
{Sod}, G.~A. 1978, Journal of Computational Physics, 27, 1

\bibitem[{Springel(2010)}]{Springel2010}
Springel, V. 2010, Monthly Notices of the Royal Astronomical Society, 401, 791

\bibitem[{Suresh \& Huynh(1997)}]{suresh_1997_amp}
Suresh, A. \& Huynh, H.~T. 1997, Journal of Computational Physics, 136, 83

\bibitem[{Taylor {et~al.}(2007)Taylor, Wu, \& Mart{\'{i}}n}]{taylor_2007_one}
Taylor, E.~M., Wu, M., \& Mart{\'{i}}n, M.~P. 2007, Journal of Computational
  Physics, 223, 384

\bibitem[{Tchekhovskoy {et~al.}(2007)Tchekhovskoy, McKinney, \&
  Narayan}]{tchekhovskoy_2007_wham}
Tchekhovskoy, A., McKinney, J.~C., \& Narayan, R. 2007, Monthly Notices of the
  Royal Astronomical Society, 379, 469

\bibitem[{Toro(1999)}]{Toro99}
Toro, E.~F. 1999, Riemann Solvers and Numerical Methods for Fluid Dynamics
  (Springer-Verlag)

\bibitem[{Wilson(1972)}]{Wilson72}
Wilson, J.~R. 1972, Astrophys. J., 173, 431

\bibitem[{Woodward \& Collela(1984)}]{Woodward84}
Woodward, P. \& Collela, P. 1984, J. Comput. Phys., 54, 115

\bibitem[{Yang(2000)}]{yang_2000_oon}
Yang, D. 2000, C++ and Object-Oriented Numeric Computing for Scientists and
  Engineers (Springer-Verlag)

\bibitem[{{Zanotti} {et~al.}(2011){Zanotti}, {Roedig}, {Rezzolla}, \& {Del
  Zanna}}]{Zanotti2011}
{Zanotti}, O., {Roedig}, C., {Rezzolla}, L., \& {Del Zanna}, L. 2011, Mon. Not.
  R. Astron. Soc., 417, 2899

\bibitem[{Zhang \& MacFadyen(2006)}]{Zhang2006}
Zhang, W. \& MacFadyen, A. 2006, The Astrophysical Journal Supplement Series,
  164, 255

\bibitem[{{Zhang} {et~al.}(2009){Zhang}, {MacFadyen}, \& {Wang}}]{Zhang09}
{Zhang}, W., {MacFadyen}, A., \& {Wang}, P. 2009, Astrophys. J., 692, L40

\bibitem[{Zrake \& MacFadyen(2012)}]{Zrake2011a}
Zrake, J. \& MacFadyen, A.~I. 2012, Astrophys. J., 744, 32

\end{thebibliography}
\end{document}